\documentclass[final,3p,times,authoryear]{elsarticle}

\usepackage{amssymb}
\usepackage{amsmath}
\usepackage[dvipsnames]{xcolor}
\usepackage{hyperref}
\usepackage{subcaption}
\usepackage{ulem}
\usepackage[T1]{fontenc}

\usepackage{lineno}

\newcommand{\kms}{km~s$^{-1} \,$}
\newcommand{\gcm}{g~cm$^{-3} \,$}
\newcommand{\cmcubed}{cm$^{-3} \,$}
\newcommand{\htwoo}{H$_2$O}
\newcommand{\cotwo}{CO$_2$}
\newcommand{\HCNconc}{[HCN] }
\newcommand{\OHconc}{$\left[ \text{OH} \right] \;$}

\journal{Icarus}

\begin{document}

\defcitealias{PierazzoChyba1999:aminosurvival}{PC99}
\defcitealias{ToddOberg2020:HCNcomet}{T\"{O}20}

\begin{frontmatter}



\title{Constraining the survival of HCN during cometary impacts}

\author[IoA]{Catriona H. McDonald \corref{cor1}}
\cortext[cor1]{Corresponding author}
\ead{catriona.mcdonald@ast.cam.ac.uk}
\author[IoA]{Amy Bonsor}
\author[Edin]{Auriol S. P. Rae}
\author[Cav]{Paul B. Rimmer}
\author[IoA]{Richard J. Anslow}
\author[Wisc_Chem,Wisc_Astro]{Zoe R. Todd}

\affiliation[IoA]{organization={Institute of Astronomy, University of Cambridge},
            addressline={Madingley Road}, 
            city={Cambridge},
            postcode={CB3 0HA},
            country={UK}}
\affiliation[Edin]{organization={School of Geosciences, Grant Institute},
            addressline={University of Edinburgh}, 
            city={Edinburgh},
            postcode={EH9 3FE},
            country={UK}}
\affiliation[Cav]{organization={Astrophysics Group, Cavendish Laboratory, University of Cambridge},
            addressline={JJ Thomson Avenue}, 
            city={Cambridge},
            postcode={CB3 0HE},
            country={UK}}
\affiliation[Wisc_Chem]{organization={Department of Chemistry, University of Wisconsin-Madison},
            city={Madison},
            state={Wisconsin},
            country={USA}}
\affiliation[Wisc_Astro]{organization={Department of Astronomy, University of Wisconsin-Madison},
            city={Madison},
            state={Wisconsin},
            country={USA}}

\begin{abstract}
Cometary impacts have been invoked as an atmosphere-independent method of stockpiling hydrogen cyanide (HCN), a key prebiotic feedstock molecule, into environments favourable for the onset of prebiotic chemistry on the early Earth. 
This work revisits the prospects for cometary delivery of HCN through new impacts simulations of idealised cometary bodies using the shock physics code iSALE combined with simple chemical modelling.
Using temperature and pressure profiles for material within spherical, non-porous comets with a high resolution of Lagrangian tracer particles, we assess the survival rate of HCN across a range of impact velocities, sizes and angles, assuming both steady state and equilibrium chemistry. 
We find that HCN survival is extremely limited at impact velocities above the escape velocity of the Earth, unless the impact occurs at extreme obliquity ($\theta \sim 15^\circ$).
We present a parametrisation of the survival of HCN as a function of impact velocity, angle, and cometary diameter, which provides an upper limit to survival in more realistic scenarios to aid with future studies investigating the role of comets in the origins of life. 
Although successful HCN delivery may be possible in our idealised model, we neglect to consider the effect of atmospheric passage and our results suggest that delivery alone is not likely to be sufficient for the onset of prebiotic chemistry. 
\end{abstract}



\end{frontmatter}


\section{Introduction}
\label{sec:intro}

The origins of life on Earth must have occurred in an environment favourable to the onset of a chemical pathway to life. 
Such a chemical pathway relies on the availability of prebiotic feedstock molecules which can form fundamental building blocks vital for life to emerge \citep{Sasselov2020:Origins}. 
A key question is how prebiotic feedstocks can be concentrated within favourable environments for prebiotic chemistry to occur.

There are numerous processes which can synthesise prebiotic molecules within the reducing atmosphere of the early Earth \citep{CatlingZahnle2020:ArcheanAtmosphere}, including atmospheric interactions with energetic solar particles \citep{Kobayashi2023:AminoAcidsSEPs}, atmospheric photochemical networks \citep[e.g.][]{Zahnle1986:AtmosphereHCN, Tian2011:AtmosphereHCN}, lightning \citep{ChameidesWalker1981:LightningHCN, Ferus2017:LightningSynth, Barth2023:Lightning}, and synthesis during meteorite atmospheric passage \citep{Chyba1992:EndoExoProd, McKayBorucki1997:ImpactSynthesis}. 
However, our knowledge of the atmosphere of the early Earth is gained by proxy through the analysis of rock samples and thus is heavily uncertain \citep{CatlingZahnle2020:ArcheanAtmosphere}.
Given in situ synthesis of prebiotically relevant molecules is dependent on unknown atmospheric conditions, exogenous reservoirs of prebiotic feedstocks may have been important for the origins of life.

Prebiotic feedstocks have been widely observed across the Solar System's small body population.
High levels of organic materials have been observed in comets through spectroscopic analyses \citep{BockeleeMorvan2004:CometsIIVolatiles, MummaCharnley2011:CometChemRev, Lippi2021:CometSpectro} and in situ analysis of cometary material by missions such as Vega \citep{KruegerKissel1987:HalleyPUMA} and Rosetta \citep[e.g.][]{LeRoy2015:Rosetta67P, Altwegg2017:Organics67P}.
Samples returned from comet 81P/Wild by the Stardust mission \citep[e.g.][]{Sandford2006:StardustOrganics, Cody2008a:StardustOrganics, deGregorio2011:StardustOrganics}, indicate the organic molecules present in the comet are rich in both oxygen and nitrogen.
Recent analyses of chondritic material returned from the asteroid (101955) Bennu \citep{Glavin2025:Bennu, McCoy2025:Bennu}, find evidence of ancient brines and a large suite of prebiotic molecules, including 14 of the 20 amino acids used in biology and all five nucleobases found in DNA and RNA.
Similarly, samples from the carbonaceous asteroid (162173) Ryugu show 15 amino acids and an array of other organic molecules \citep[e.g.][]{Naraoka2023:RyuguOM}.

This inventory of organic matter present in small bodies has also been found to survive delivery to the Earth's surface.
The Murchison meteorite which fell in Australia in 1969 contains a large amount of amino acids and other organic molecules \citep{Kvenvolden1970:Murchison, Cronin1988:MurchisonReview}, with 70~\% of the meteorite's carbon content being insoluble organic matter \citep{SchimttKopplin2010:Murchison}.
A number of carbonaceous meteorites recovered from the Antarctic show evidence of amino acids and potentially membrane forming lipids \citep{Shimoyama1986:CCAminoAcids, Shimoyama1989:CCAminoAcids, Naraoka1999:CCAminoAcids}.
Additionally ultra-carbonaceous Antarctic micrometeorites (UCAMMs) rich in organic matter show very high N/C ratios compared to the insoluble organic matter found in other carbonaceous meteorites \citep{Dartois2013:AntarcticMicrometeorites} and can survive large scale melting and volatile loss during atmospheric passage at velocities $< 15$~\kms \citep{Kohout2014:CosmcDust}. 
About $25$~\% of the flux of unaltered carbon accreting onto the Earth's surface is in the form of UCAMMs \citep{Rojas2021:MicrometeoriteFlux}.

Extensive cratering observed across the terrestrial planets indicates that they underwent a period of sustained bombardment from planetesimals after the formation of their crusts \citep{Wetherill1975:LHB0, BottkeNorman2017:Review_LHB, Nesvorny2023:LunarBombard}. 
Although the duration, timing, and dynamical processes driving the bombardment are contested \citep[e.g.][]{Ryder1990:LunarBombard, Bottke2012:ArchaeanBombard, Cohen2023:ReviewMoonImpact}, it is indisputable that a large number of bodies which could have harboured prebiotic feedstocks impacted the rocky planets early in their evolution. 
The earliest direct evidence for complex life on Earth, in the form of microfossils permineralized in the Western Australian Apex chert, dates to approximately $\sim 3.5$~billion years ago, and carbon isotope measurements of even older rocks suggest that precursor biotic processes occurred $\sim 3.8$~billion years ago \citep{Mojzsis1996:EarlyLife}.
Thus, the appearance of biotic activity on the early Earth occurs shortly after this influx of impactors.

\cite{Clark1988:CometPond} suggested that a fortuitous soft comet impact could produce a `procreative' pond of cometary melt material including organics. 
The varying climate on early Earth could cause evaporative processes and wet-dry cycling, of such a pond, which can store volatile molecules in salts such as ferrocyanide and promote polymerization aiding in the formation of RNA \citep{Pearce2017:RNAPond, Campbell2019:WetDryCycles, Sasselov2020:Origins}.
Comet ponds may then have played a vital role in the origins of life on early Earth.  
The idea that extraterrestrial material, in the form of comets, asteroids, or interplanetary dust particles, falling onto the surface of the early Earth could deliver prebiotic feedstocks was first proposed over a century ago \citep{ChamberlinChamberlin1908:TerrestrialOrganics} and has been continually considered since \citep[e.g.][]{LederbergCowie1958:Moondust, Chyba1990:ETAminoAcids, ToddOberg2020:HCNcomet, Walton2024:IDPs}.

Consequently, a large number of previous studies have investigated the potential for impacts of extraterrestrial material to deliver organic materials to the Earth. 
Shock experiments have highlighted that extreme temperatures and pressures involved with even low velocity impact processes can lead to the destruction of organic molecules \citep[e.g][]{Blank2001:AqueousShockChemistry, Bertrand2009:AminoSurvivMeteor, Burchell2009:OrganicsAerogel, Burchell2014:OrganicsHypervelocityImpacts}.
Experiments exposing organic-rich meteoritic material to high temperatures emphasise how the survival of organic molecules sensitively depends on the thermal history of the impactor \citep{Cody2008b:ChondriteThermos, Kohout2014:CosmcDust}. 
Finally, numerical modelling of both the impact and chemical processes involved with extraterrestrial delivery, which exceed the conditions experiments can reasonably reproduce, suggest that the survival of organic matter would be extremely limited in large impactor delivery events \citep[e.g.][]{Chyba1990:CometaryDelivery, Ross2006:CometImpactChem}.

\cite{PierazzoChyba1999:aminosurvival} (hereafter \citetalias{PierazzoChyba1999:aminosurvival}) used hydrodynamic simulations to model the survivability of amino acids inside a large, icy, cometary body impacting onto a planet with an ocean layer.
Their simulations were restricted to vertical impacts, and utilised scaling relationships to extend their results to oblique impacts, which make up the vast majority of actual impacts \citep{Gilbert1893:LunarFace, Shoemaker1962, PierazzoMelosh2000a:ObliqueReview}.
They found that larger impact velocities, larger impactor size, and smaller impact obliquities all decreased the likelihood of amino acid survival, with peak survival rates of $\sim 10$~\% in best case scenarios.

\cite{ToddOberg2020:HCNcomet} (hereafter \citetalias{ToddOberg2020:HCNcomet}), extended the work in \citetalias{PierazzoChyba1999:aminosurvival} to consider the survival of HCN during impacts, by parametrising the impact temperature and pressure profiles presented in \citetalias{PierazzoChyba1999:aminosurvival} and applying a time-dependent chemical model. 
They find that HCN survival is similarly limited by the impactor's size, velocity and angle, but that cometary impacts may provide a feasible, atmosphere-independent way to provide a localised enhancement in HCN.

With the goal of further exploring the survival of prebiotic molecules during cometary impacts, Section~\ref{sec:chemistry} introduces, and motivates, a simple chemical model we use to determine the destruction of HCN during an impact.
Section~\ref{subsec:impact_2D} compares new two-dimensional simplified cometary impact simulations with the results of \citetalias{PierazzoChyba1999:aminosurvival} and \citetalias{ToddOberg2020:HCNcomet}. 
Section~\ref{subsec:impact_3D} introduces three-dimensional impact simulations to consider the survival of prebiotic feedstocks in oblique cometary impacts on the early Earth. 
In Section~\ref{sec:discussion} we discuss the implications of our results and the consequences for the origins of life. 
Finally, we summarise in Section~\ref{sec:conc}.

\section{Prebiotic Chemistry from Comets} \label{sec:chemistry}
Alongside the direct discovery of multiple amino acids in meteorites, volatile-rich molecules are observed extensively in cometary bodies \citep[see][for a recent review]{McKayRoth2021:CometOrganics}.
Spectroscopic surveys have detected the presence of volatile-rich molecules such as acetylene (C$_2$H$_2$), methanol (CH$_3$OH), ammonia (NH$_3$) and hydrogen sulfide (H$_2$S) at the percent level relative to water in at least ten comets \citep{Crovisier2006:Comets, MummaCharnley2011:CometChemRev}.
The delivery of these molecules during impacts could provide vital ingredients needed to initialise prebiotic chemistry.

Hydrogen cyanide (HCN) has been extensively explored as a vital prebiotic feedstock for the origins of Earth's life and is present in comets at around $0.1$~\% relative to water \citep[e.g.][]{MummaCharnley2011:CometChemRev}. 
\cite{Oro1960:AdenineSynthesis} and \cite{OroKimball1961:Purines} showed that the RNA and DNA nucleobase adenine can be formed through the polymerisation of molecules such as ammonium cyanide and HCN. 
\cite{Miller1957:AminoAcids} famously identified the synthesis of amino acids through  electric discharges,  with HCN being a vital intermediary step to producing  prebiotic molecules through Strecker synthesis \citep[e.g.][]{Singh2022:Strecker}.
With formamide as an intermediary, HCN can lead to the formation of purine nucleobases \citep{Sanchez1966:Purine, Sanchez1967:PrebioticSynthesis, Saladino2015:MeteoritesNucleosides, Saladino2016:MeteoritesFormamide}.
The `RNA world' hypothesis for the origins of life also invokes HCN in the formation of purine RNA precursors \citep{Becker2016:PurineNucleoside, Benner2020:RNA}.
Finally, \cite{Patel2015:cyanosulfidic} describes a cyanosulfidic protometabolism where the reductive homologation of HCN and its derivatives under UV light produces the precursors of ribonucleotides, amino acids, and lipids. 

Both \cite{Patel2015:cyanosulfidic} and \cite{Sasselov2020:Origins} outline similar geochemical scenarios through which the cyanosulfidic pathway to life can proceed. 
These scenarios rely on HCN, either delivered, or produced, during high velocity impacts dissolving in surface water and interacting with available ferrous iron to form ferrocyanide \citep[e.g.][]{TonerCatling2019:Ferrocyanide}.
Episodic drying of the surface water driven by climatic variations could then create an evaporite layer containing ferrocyanide salts.
Thermal metamorphosis of this evaporite layer driven by geothermal activity, such as volcanic activity or large impacts, can then produce cyanide, cyanamide, and carbide salts. 
Through the reintroduction of water to this evaporite layer, and exposure to ultraviolet light, the cyanosulfidic chemical pathway can proceed \citep{Patel2015:cyanosulfidic, Xu2018:HCN, Rimmer2021:Photochemistry}. 

There are reasons why HCN is invoked in so many prebiotic scenarios.
It looks like a fragment of the nitrogenous bases life uses to build RNA and DNA, not just in terms of atomic order (H-C$\equiv$N), but also in its electrons.
The carbon has an oxidation state of $+2$, and so is only two electrons away from the oxidation state for simple sugars and the nitrogen has an oxidation state of $-3$, the same oxidation state as nitrogen in virtually all organic molecules life uses \citep[see Fig. 1 of][]{Sasselov2020:Origins}.
The C$\equiv$N triple bond contains $\sim 2.8$ eV of energy, $\sim 10$ times the energy released by converting adenosine triphosphate (ATP) to adenosine diphosphate (ADP) \citep{Giesen1980:Phosphorylation}. 
This energy is accessible selectively, through a few productive chemical pathways \citep{Green2021:UVPhotochem}, and in some environments only slowly degrades into formamide, and then formate \citep[][Murali \& Rimmer, in prep.]{Miyakawa2002:ColdOrigins}.
Although this form of chemical storage is very different from the chemical storage modern life uses (e.g. ATP) it is similar to energy storage of modern life in terms of its atomic structure, electronic structure and stored energy.
Thus HCN is an incredibly useful, and versatile, prebiotic feedstock.

\citetalias{ToddOberg2020:HCNcomet} have previously considered the prospects for delivering HCN during cometary impacts by applying chemical networks modelling the destruction of HCN to the parametrised results of \citetalias{PierazzoChyba1999:aminosurvival} (further discussed in Section~\ref{subsec:impact_2D}).
Their chemical network contains 31 species and 111 reactions and time dependently models the thermal degradation of volatiles included (\htwoo, \cotwo, CO and HCN), HCN degradation by radical reactions, and nitrile chemistry.
As HCN is the only nitrogen carrier in their model, they neglect the impact driven generation of HCN.
At each time step, the temperature dependent reaction rate constants are used to calculate the concentration of each species, with concentrations updated to account for the expanding volume during impact. 
In addition to the full chemical network discussed above, they also considered a smaller network which only included thermal \htwoo~decomposition and the OH driven destruction of HCN. 
Applying their models to the temperature and pressure profiles of the \citetalias{PierazzoChyba1999:aminosurvival} impact simulations, they find that the simplified chemical network overestimates the levels of HCN degradation compared to the full model by only a small amount. 

Thus, although a more complete chemical network may allow us to better understand the degradation of HCN, in the following work we elect to use a similar simple model to determine upper limits on the survival of HCN during cometary impacts.
The inclusion of more realistic chemical compositions for the comets, and complex chemical networks for the impact will be reserved for future work. 

Our simplified chemical network considers only the two following reactions 
\begin{equation}
    \text{H}_2 \text{O} \leftrightharpoons \text{H} + \text{OH},
    \label{eq:thermalWaterDecomp}
\end{equation}
\begin{equation}
    \text{HCN} + \text{OH} \rightarrow \text{products},
    \label{eq:HCNdegrad}
\end{equation}
where we are completely agnostic about the precise products of the HCN degradation in Equation~\ref{eq:HCNdegrad} and only interested in the loss of the HCN. 
Thus, the time, and temperature, dependent number density of HCN (\HCNconc cm$^{-3}$) can be described by the following equation
\begin{equation}
    \left[ \text{HCN}\right](t) = \left[ \text{HCN}\right]_0 \, \text{exp} \, \{ -k \left[ \text{OH} \right] t \},
    \label{eq:HCN_degrate}
\end{equation}
where \OHconc (cm$^{-3}$) is the number density of OH and $k$ is the reaction rate constant for Equation~\ref{eq:HCNdegrad} as identified from \url{kinetics.nist.gov} \citep{Atkinson2001:HCNreactrate} and given by 
\begin{equation}
    k = 1.2 \times 10^{-13} \, \text{cm}^3 \, \text{s}^{-1} \;\text{exp} \left\{ \frac{-400.24 \, \text{K}}{T} \right\},
    \label{eq:k_HCN}
\end{equation}
where $T$ is the temperature in K.
Choosing to consider the product agnostic degradation of HCN, likely fails to capture the true rate of HCN decay and thus our estimate of \HCNconc is conservative.

Given the extreme temperatures and pressures involved with impact processes, and the degradation of HCN in our simple model being heavily dependent on \OHconc, we calculate this number density under two assumptions.
Firstly, we calculate the temperature dependent \OHconc by assuming the reaction in Equation~\ref{eq:thermalWaterDecomp}, and hence \OHconc, is in steady state, using the following forward ($k_f$) and backwards ($k_b$) reaction rates identified from \url{kinetics.nist.gov} \citep{Baulch1992:NIST_data},
\begin{equation}
    k_f = 1.4 \times 10^{11} \, \text{s}^{-1} \; \text{exp} \left\{ \frac{-52800 \, \text{K}}{T} \right\},
    \label{eq:k_forward}
\end{equation}
\begin{equation}
    k_b = 1.91 \times 10^{-10} \, \text{cm}^3 \, \text{s}^{-1} \left( \frac{T}{298 \, \text{K}} \right).
    \label{eq:k_backwards}
\end{equation}
Secondly, we calculate the equilibrium concentration of OH using the open source code \texttt{FastChem} \citep{Stock2018:FastChem},  which calculates chemical equilibrium compositions for given temperatures and pressures.

\cite{MummaCharnley2011:CometChemRev} report the spectroscopic identification of HCN in more than 10 comets at $\sim 0.1$~\% abundance relative to water. 
Thus, following \citetalias{ToddOberg2020:HCNcomet} we assume our comets contain the molecules \htwoo : HCN in the mixture 100 : 0.15.
Assuming an idealised comet that is purely spherical, has zero porosity\footnote{We acknowledge here that real comets are not spherical, nor are they non-porous, and neglecting to consider these aspects significantly simplifies our simulation scheme at the expense of accuracy. 
These assumptions are chosen to create a `best-case' scenario for survival with an idealised comet that minimises impact temperatures and pressures and increases survival, thus allowing us to place an upper limit on survival.
We further discuss the limitations these choices place on our models in Section~\ref{sssec:limit_comet}.}, and that the HCN is uniformly distributed throughout the comet allows us to calculate the initial concentration of HCN in the body. 
As an example, for a 1km diameter comet with an average cometary density of $1$~\gcm, \HCNconc$_0 \approx 5.00 \times 10^{19}$~\cmcubed.  

\begin{figure}
    \centering
    \includegraphics[alt={This figure shows how long it takes 99 percent of a sample of HCN to degrade as a function of temperature. For all the data this critical time decreases rapidly with increasing temperature. At 4000 Kelvin, the curves for equilibrium chemistry jump downwards and become steeper.}, width=0.7\textwidth]{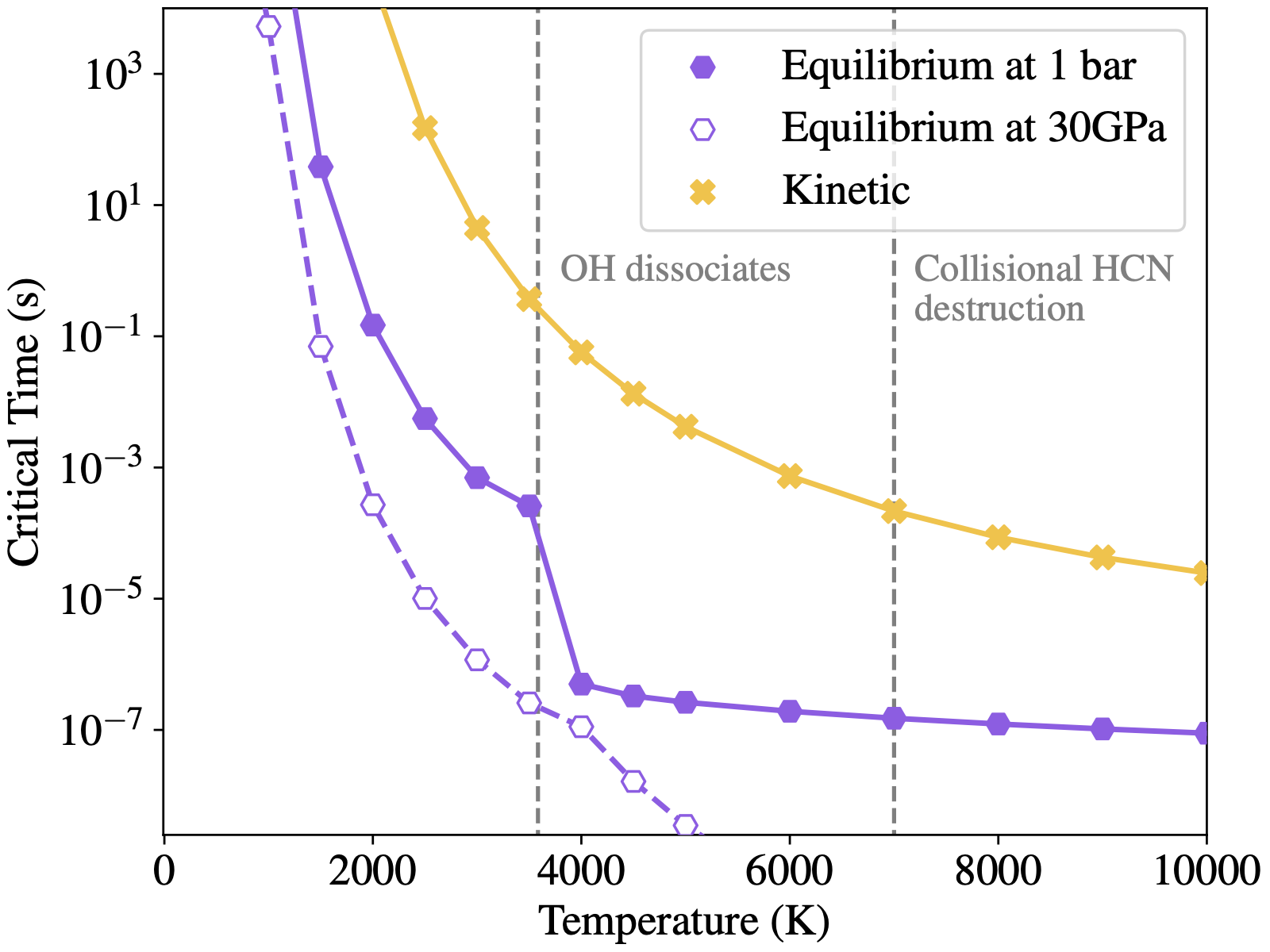}
    \caption{The critical time for a concentration of HCN to be reduced by $99$~\%, assuming it is uniformly heated to the given temperature.
    In filled purple octagons we show the critical time assuming [OH] is in equilibrium at $1$~bar, unfilled purple octagons assume [OH] is in equilibrium at $30$~GPa (the approximate peak pressure generated by an impact at $5$~\kms), and in yellow crosses we show the critical time assuming [OH] is in kinetic steady state, as described in the legend.
    The dashed grey line at $T \sim 3600$~K shows the approximate temperature where OH itself dissociates, at this point the equilibrium curve begins to follow HCN destruction by atomic oxygen, a quicker reaction. 
    The dashed grey line at $T \sim 7000$~K identifies the approximate temperature above which HCN molecules have enough kinetic energy to undergo collisional destruction. 
    Above this temperature, all curves likely underestimate the destruction rate of HCN since we neglect collisional destruction.
    For both chemistries, at temperatures above $\sim 4000$~K the majority of HCN will be destroyed in just a fraction of a second. 
    In reality, the true critical time for 99~\% of a concentration of HCN to be destroyed likely lies closer to our equilibrium curve as described further in the text.}
    \label{fig:critHCNconc}
\end{figure}
We provide an initial estimate for the survival of HCN as given by Equation~\ref{eq:HCNdegrad} in Figure~\ref{fig:critHCNconc} which shows the time for \HCNconc to be reduced to $1$~\% of its initial value, assuming it is uniformly heated to a range of temperatures under both equilibrium and steady state kinetic chemistry.
We show two example equilibrium chemistries assuming constant pressures of $1$~bar (current atmospheric pressure of Earth) and $30$~GPa (approximate peak pressures during a $5$~\kms impact), highlighting that high pressures expedite HCN destruction in our simple model, especially at higher temperatures.
This is consistent with the results of shock experiments testing the survival of organic molecules under varying pressures which find that even at relatively low pressures ($< 5$~GPa) survival of organics is extremely limited \citep[e.g.][]{Peterson1997:AminoAcidShocks, Burchell2014:OrganicsHypervelocityImpacts}. 
\cite{Blank2001:AqueousShockChemistry} find that high pressures can reduce the rate of pyrolysis for organic molecules, suggesting that simultaneous temperatures of $7000$~K and $100$GPa (approximately the conditions caused by a vertical impact of a water ice comet against basalt at escape velocity) would have a similar rate of thermal degradation to material $900$~K and $\sim1$~bar. 
However, the period of impact induced high pressure is incredibly short (see Figure~\ref{fig:TPSurvPaths}), likely shorter than the timescale for HCN destruction at the equivalent temperatures and pressures discussed above, thus this shielding process is not effective in large impacts and sustained high temperatures will still drive degradation of organic molecules \citep{Ross2006:CometImpactChem}.

At the approximate temperature at which OH dissociates ($T \sim 3600$~K), our equilibrium calculation swaps to considering the destruction of HCN by atomic O, with the following rate constant \citep[\url{kinetics.nist.gov},][]{Baulch1992:NIST_data}, 
\begin{equation}
    k_\text{O} = 3.60 \times 10^{-13} \, \text{cm}^{3} \, \text{s}^{-1} \left( \frac{T}{298 \, \text{K}} \right)^{2.10} \text{exp} \left\{ \frac{-3080.17 \, \text{K}}{T} \right\}.
    \label{eq:k_O}
\end{equation}
This reaction leads to a steeper decay rate for \HCNconc as shown by the deviation in the trajectory of the equilibrium model at 1 bar and implies that survival of HCN above $\sim 3600$~K is minimal for all chemistries.

HCN molecules themselves have enough kinetic energy to be destroyed through collisions at $T \sim 7000$~K. 
As our calculations neglect to consider collisional destruction, the kinetic chemistry model likely overestimates the critical time to destroy 99~\% of HCN. 
Thus, we expect no HCN to survive at temperatures above $\sim 7000$~K, and at the point where OH thermally dissociates ($\sim 3600$~K) 99\% of a sample of HCN will degrade in $1.5 \times 10^{-4} - 0.15$ seconds, whereas HCN may survive for tens of seconds at temperatures below $2000$~K.
The duration of our impact simulations is dependent on the impactor size and velocity (as further described in Section~\ref{sec:impacts}), but is typically on the order of $0.1-1$~s.
Full kinetic modelling suggests that OH and other radicals reach equilibrium quickly at $T > 2000$~K, and so more detailed, realistic, kinetic modelling will likely lead to results closer to the equilibrium curve in Figure~\ref{fig:critHCNconc} than the steady state kinetics curve consistent with previous work comparing kinetics to equilibrium \citep{Liggins2023:KineticVolcanicAtmospheres}. 

\section{Impact Simulations} \label{sec:impacts}
We next produce a suite of impact simulations to constrain the survival of HCN in a range of scenarios. 
We first consider two dimensional simulations to provide direct comparisons with the work of \citetalias{PierazzoChyba1999:aminosurvival} and \citetalias{ToddOberg2020:HCNcomet}.
We carry out these simulations using the iSALE-2D shock physics code \citep{Wunnemann2006:iSALE}, which is based on the SALE hydrocode solution algorithm \citep{Amsden1980:SALE}.
Several modifications and extensions to iSALE-2D have been made, including adding an elasto-plastic constitutive model, fragmentation models, various equations of state (EoS), and multiple materials \citep{Melosh1992:ImpactFrag, Ivanov1997:HydroStrength}.
Further developments of the code, not utilised in this study, are outlined by \cite{Collins2004:iSALE, Wunnemann2006:iSALE, Collins2011:iSALECompact, Collins2014:iSALEDilatency}.

We also produce simulations of oblique impacts using iSALE-3D \citep{Elbeshausen2009:iSALE3D} (discussed in Section~\ref{subsec:impact_3D}) and a solver as described by \cite{Hirt1974:LEflowspeeds}.
The development history of iSALE-3D is described by \cite{Elbeshausen2011:iSALE3D}, and the code benefits from similar modifications and extensions to iSALE-2D.
Both iSALE-2D and iSALE-3D have been benchmarked against other hydrocodes \citep{Pierazzo2008:ImpactValidation} and validated against experimental data from laboratory scale impacts \citep{Pierazzo2008:ImpactValidation, Davison2011:ObliqueImpacts, Miljkovic2012:PorousImpacts}.

Throughout all of our impact simulations, we assume that the cometary projectile is made of non-porous pure water ice (over which we enforce a homogeneous distribution of HCN, as discussed in Section~\ref{sec:chemistry}), uniformly at $200$~K prior to the impact. 
We further assume that our projectiles are perfectly spherical, with iSALE-2D representing the projectiles in a cylindrically symmetric grid for computational efficiency.
To approximate the surface of early Earth, we use a single layer of basalt with a surface temperature of $293$~K, and assumed standard planet radius and gravitational field strength values.
We use the analytic equation of state program (ANEOS; \cite{ThompsonLauson1974:ANEOS, Melosh2007:ANEOS_SiO2}) to generate the equations of state for both basalt and ice in our simulations, and utilise a purely hydrodynamic model which represents the materials as inviscid fluids with no strength.
Although a strength model would be important for considering the effect of the impact on the target surface \citep{KurosawaGenda2018:PlasticDeformation}, we are only concerned with the pure water ice impactor material which will vaporize even at low impact velocities, thus can be well approximated as a fluid.
The implications of including a strength model for our impactor is discussed in Section~\ref{sssec:limit_comet}.
These simplifying assumptions create an idealised comet which undergoes impacts most probable for HCN survival (with low temperatures and pressures) allowing us to place an upper limit on survival.

Our simulations run just long enough for the shock wave to move upwards through the projectile and generate a release wave at the top surface that travels back through the projectile, unloading it from the shocked state.
In order to take into account the effect of impactor diameter ($d$) and impact velocity ($v_\text{imp}$) on the duration of the shock pulse we introduce a time scaling parameter given as follows $t_\text{scale} = d/(4v_\text{imp})$ \citep{Collins2016:iSALEDellenManual}. 
The interval between each data saving cycle is then defined as $t_\text{save} = 0.05 t_\text{scale}$ and the final simulation time is set to $t_\text{sim} = 6 t_\text{scale}$.

The rarefraction wave releases the cometary material from its shocked state and hence ends the period of extremely high temperatures and pressures. 
Although the impactor material will retain elevated temperatures beyond this point, the peak temperatures and pressures, which would have the largest effect on the survival of HCN, will have passed upon transmission of the rarefaction wave \citep[e.g.][see Figure~\ref{fig:TPSurvPaths}]{Melosh1989:ImpactCratering}. 

The resolution of grid-based impact simulations, such as iSALE, is conventionally described in cells per projectile radius (CPPR), with higher CPPR providing higher accuracy results. 
Although we note that the accuracy of an impact simulation is not only dependent on the CPPR but also the efficiency of heating in the impact through parameters such as porosity and material strength \citep{Davison2010:PorousImpacts}.
\begin{figure*}[t!]
    \centering
    \begin{subfigure}[t]{0.45\textwidth}
        \caption{}
        \includegraphics[alt={The mass fraction of a body exceeding 20000 Kelvin linearly decreases with increasing cells per projectile radius.}, width=0.88\textwidth]{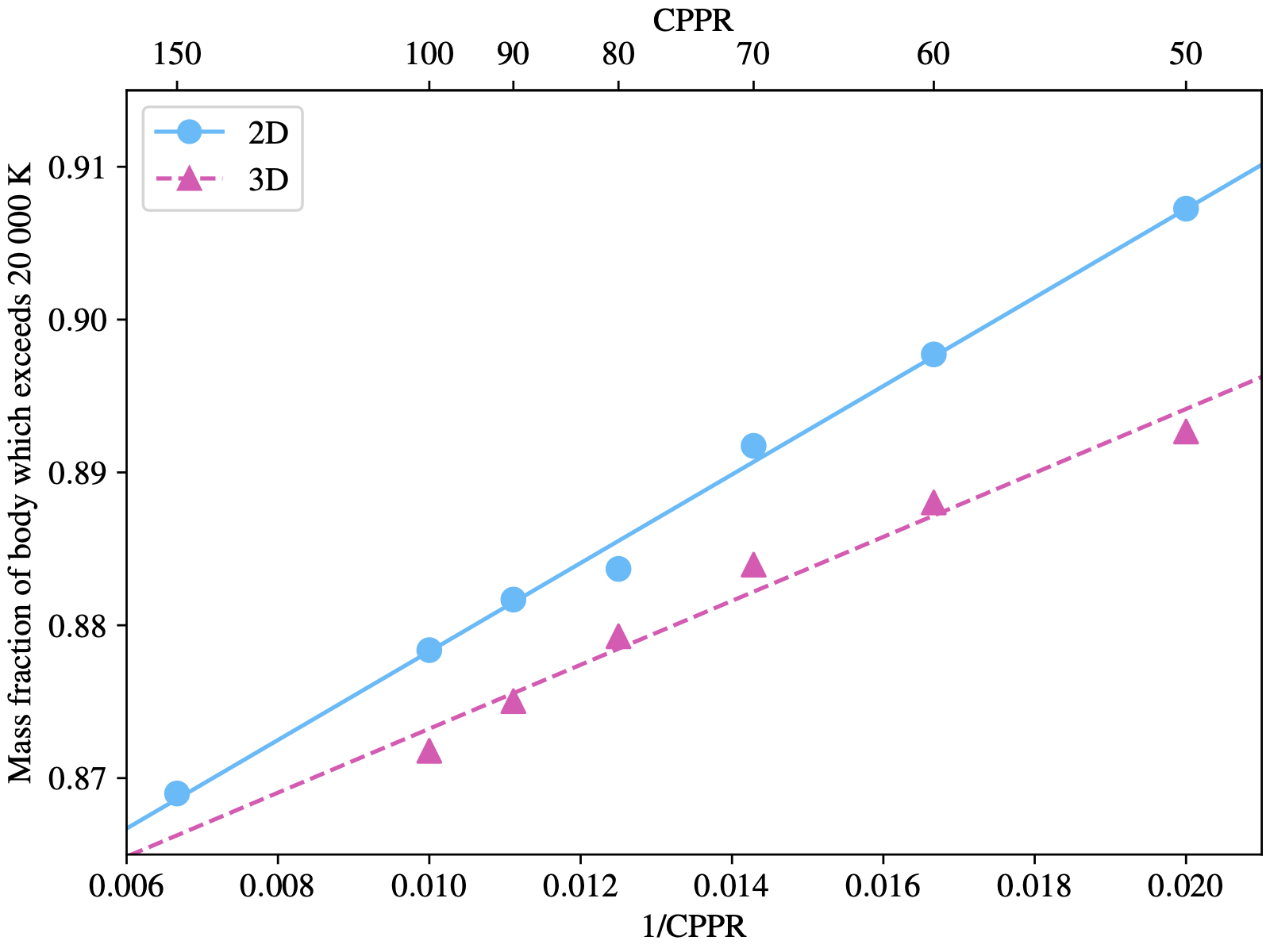}
        \label{subfig:melt_res}
    \end{subfigure}
    \begin{subfigure}[t]{0.455\textwidth}
        \caption{}
        \includegraphics[alt={The percentage of HCN which survives an impact is approximately constant with increasing cells per projectile radius. There is an approximate 2 per cent difference between the surviving HCN for 2D and 3D simulations, with 2D simulatins being lower.}, width=0.9\textwidth]{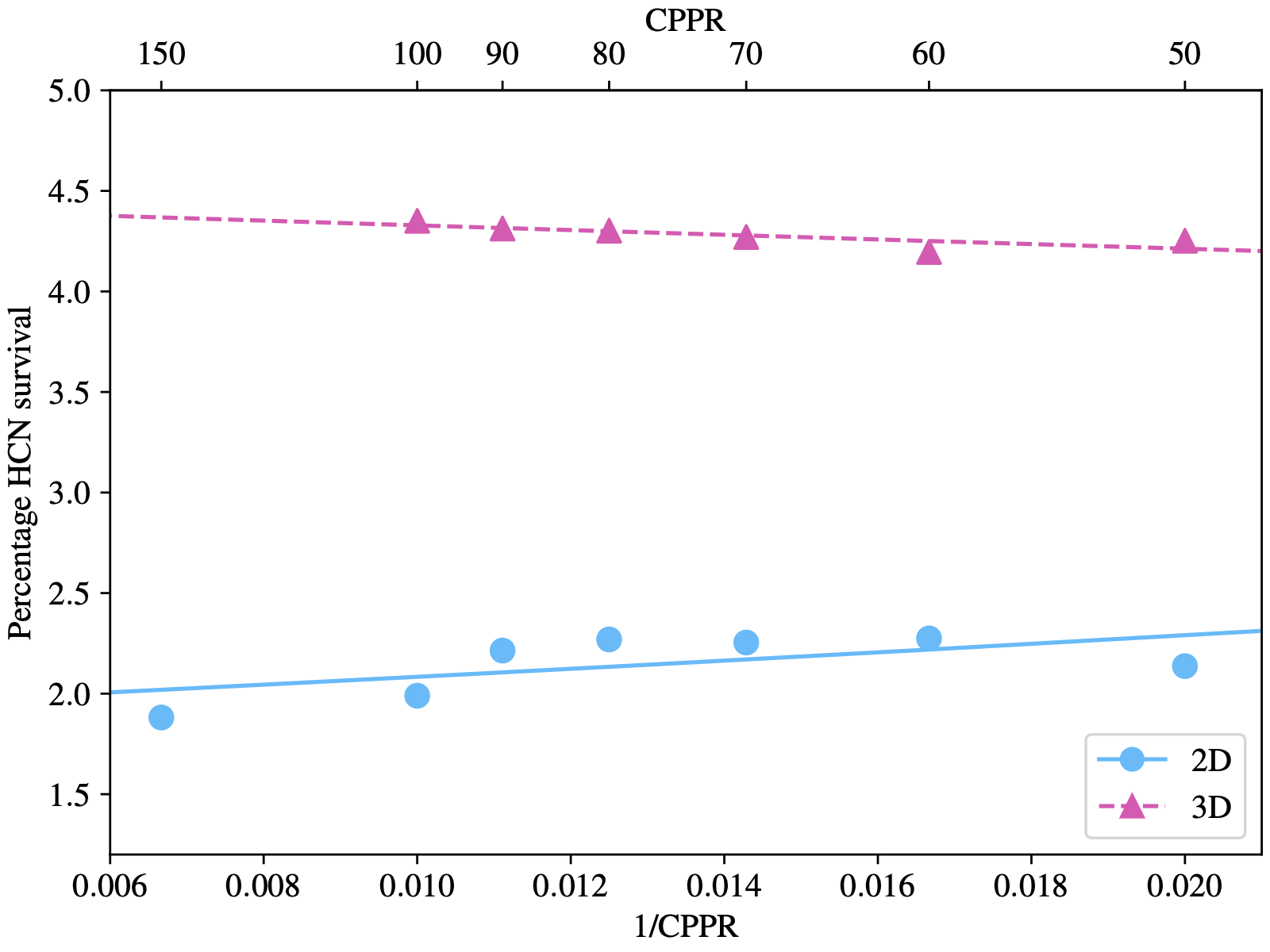}
        \label{subfig:chem_res}
        \end{subfigure}
    \caption{(a) The mass fraction of a $1$~km diameter comet's body which exceeds $20 \, 000$~K during an impact at $30$~\kms for a range of CPPR.
    Results from 2D simulations are shown by blue circles and 3D results in pink triangles in both panels. 
    As the resolution of the simulations increase, the fraction of the body exceeding $20 \, 000$~K decreases. 
    A linear fit to both sets of data are shown by solid (2D) and dashed (3D) lines as also shown in the legend, both fits converge to a true mass fraction (ie, at infinite resolution) of a $1$~km which exceeds $20 \, 000$~K of $\sim 0.85$.
    Thus, 2D simulations at CPPR = 50 may overestimate the `true' mass fraction of a comet exceeding a given temperature by around $5$~\%, and 3D simulations may overestimate the mass fraction by $\sim 4$~\%.
    (b) The total percentage of HCN surviving a $1$~km diameter cometary impact at $10$~\kms for a range of CPPR.
    The survival percentages are approximately consistent across all resolutions with for both 2D and 3D simulations, the higher survival calculated for 3D is due to higher survival rates towards the back of our 3D simulations which are not captured in the 2D cases. 
    A similar straight line fit to panel a suggests a `true' surviving percentage of 1.90~\% for 2D models and $4.45$~\% for 3D.
    }
    \label{fig:resolution_tests}
\end{figure*}

Figure~\ref{subfig:melt_res} shows the mass fraction of a $1$~km diameter comet that exceeds $20 \, 000$~K during an impact at $30$~\kms for a range of different CPPR in both 2D and 3D simulations converging towards a single value at high resolution.
In these simulations, all other parameters are kept constant and the size of the simulation grid scales with the CPPR.
There is no 3D simulation with 150 CPPR as this simulation proved excessively computationally expensive. 
Following previous numerical works \citep{Wunnemann2008:PorousMelt, Davison2010:PorousImpacts}, the mass, or volume, of material exceeding a given temperature can be approximated as an exponential function of the form $M(>T) = a \, \text{exp}\left[ b/\text{CPPR}\right]$.
The true mass of the impactor which would exceed the given value is then found by extrapolating this function to $1/\text{CPPR} = 0$, ie. infinite resolution.
For both our 2D and 3D simulations this suggests the `true' mass fraction of a $1$~km diameter comet impacting at 30~\kms which exceeds $20\,000$~K is approximately $0.85$.
In our simulations, at a resolution of 100 CPPR, $87.84$~\% ($87.18$~\%) of a 2D (3D) comet impact simulation exceeds $20 \, 000$~K and this increases to only $90.73$~\% ($89.27\%$) for 2D (3D) simulations at 50 CPPR.
Thus, utilising a CPPR of 50 for our purposes should lead to differences from the true value of no more than around $6~\%$, and we adopt this CPPR throughout the rest of the work.

To confirm that our simulations are able to appropriately capture the survival of HCN as is the goal of this paper, we carry out an additional resolution test.
Figure~\ref{subfig:chem_res} shows the surviving percentage of HCN in a $1$~km diameter comet impacting at $10$~\kms for the same range of resolutions (see Section~\ref{subsec:impact_2D} for details on the survival calculation).
The surviving percentage of HCN is approximately constant for all resolutions varying by only a fraction of a percent. 
The $\sim 2$~\% difference in the survival between the 2D and 3D simulations is due to the 3D simulations being able to better encompass the effect of the comet's volume and constitutes a smaller variation than between the melt fraction volumes for 50 CPPR and 80 CPPR as in Figure~\ref{subfig:melt_res}.
Thus, utilising a CPPR of 50 should sufficiently resolve HCN survival, even in cases of limited survival.

In all of our 2D simulations we utilize a grid size of $400 \: \times \: 450$ cells (horizontal vs vertical) producing a grid $8\times$ the width of the cometary projectile.
For our 3D simulations we utilise a fixed grid size for each impact angle tested and offset the point of impact away from the centre of the simulation grid to account for the increasingly downrange motion of projectile material with increasing impact angle. 
When considering vertical 3D simulations $(\theta = 90^\circ)$ we use a grid size of $456 \: \times \: 456 \: \times 80$, but at $\theta = 15^\circ$ we use a grid of $684 \: \times \: 456 \: \times \: 100$.
These extended grid sizes prevent impactor material exceeding the limits of the simulation grid and escaping the simulation.

Although choosing to utilise a CPPR of 50, does not provide a grid resolution improvement over \citetalias{PierazzoChyba1999:aminosurvival}, one significant advancement in our simulations is the ability to record the state of a large number of Lagrangian tracer particles across the duration of the simulations. 
These massless tracer particles move through the Eulerian simulation grid following the simulated material by using the local velocity recorded at the tracer's current position \citep{Pierazzo1997:MeltProduction}.
The particles sample the Eulerian grid properties and allow us to record the temperatures and pressures of the grid material as a function of time.
iSALE contains two methods for calculating tracer motion, the first (\texttt{VELOCITY}) uses the velocities recorded at the current grid cell's nodes to interpolate the velocity of the tracer and update its position at each timestep. 
We utilise the second method (\texttt{MATERIAL}) which calculates tracer velocities based on the volumetric flux of materials in and out of the tracer's grid cell \citep{Davison2016:MesoscaleiSALE} and more closely tracers material motion.
Neither method is perfect and can lead to tracers being removed from the simulation to prevent anomalous results. 
In particular by using the \texttt{MATERIAL} method, tracer positions are checked at each time step and are removed from our simulations if they become separated from the grid material they were supposed to be tracking.
In a 2D test simulation of a $1$~km diameter comet impacting at $10$~\kms we find that $1.96$~\% of tracers are deleted while using the \texttt{MATERIAL} algorithm compared to $2.32$~\% with the \texttt{VELOCITY} method.
Thus, we choose to utilise the \texttt{MATERIAL} algorithm for this study.
\citetalias{PierazzoChyba1999:aminosurvival} distribute 100 Lagrangian tracers throughout their impactors, however, we are able to place a tracer particle in the centre of every grid cell the impactor initially occupies, totalling 3923 tracers for our 2D simulations and 259172 tracers for our 3D simulation.

In order to fully consider the prospects for delivering HCN with cometary impacts, we consider a range of impactor sizes, velocities, and angles as described in the following sections. 

\subsection{2D Simulations} \label{subsec:impact_2D}

\subsubsection{Impact velocity} \label{sssec:ImpactVelocity}
The first parameter we consider in this work is the impact velocity ($v_\text{imp}$) of the comet.
\cite{HughesWilliams2000} determine the predicted impact velocities of all numbered Solar System comets with a perihelion $< 1$~au.
They find that if the comets were to impact on Earth, those with short periods ($< 20$~yr) would have a mean impact velocity of $20$~\kms, with the tail of the distribution extending up to $70$~\kms. 
Long period comets ($P > 430$~yr) have mean impact velocities in excess of $50$~\kms.
Numerical simulations focussed on the formation of the terrestrial planets similarly find that the impact velocities of comets, asteroids, and other planetesimals on the terrestrial planets would be in excess of $20$~\kms \citep{Rickman2017:LHB_CometImpacts, Brasser2020:ImpactChronology}.

The prospects for an impacting comet to survive passage through the atmosphere of the Earth, at any impact velocity, is heavily tied to the size of the comet at the point of entry (see Section~\ref{ssubsec:2D_size}), with small comets failing to reach the surface of the Earth, instead airbursting at high altitude. 
\cite{Anslow2025:atmosentry} introduces an open source model (\texttt{atmosentry}) to model the evolution of comets during atmospheric passage. 
Their model shows that small objects which are able to survive passage through an atmosphere similar to modern Earth's will have their velocities reduced.
They show that a $150$~m radius comet which enters the atmosphere at around $22$~\kms will deposit energy in the atmosphere, lose mass, and subsequently impact the surface at around $12$~\kms. 
If instead the same comet enters the top of the atmosphere at around $11$~\kms, roughly the escape velocity of Earth, the comet will impact the ground at around $7$~\kms. 
Larger bodies ($\gtrsim 1$~km) remain relatively unaffected by the presence of the atmosphere, and therefore do not have their velocities drastically reduced.

Thus, we consider the possible impact velocities for cometary bodies on Earth to be in the range 5-50~\kms to encompass both extremes.
This wide range of possible impact velocities has a large effect on the temperatures experienced by the comet during the impact. 
\begin{figure}
    \centering
    \includegraphics[alt={For both halves of this figure, colours vary with temperature achieved during an impact, and overlaid contour lines for particular temperatures. The lower edge of the body achieves 74000 K, and the temperatures decrease moving upwards, but with raised temperatures towards the centre.}, width=0.7\textwidth]{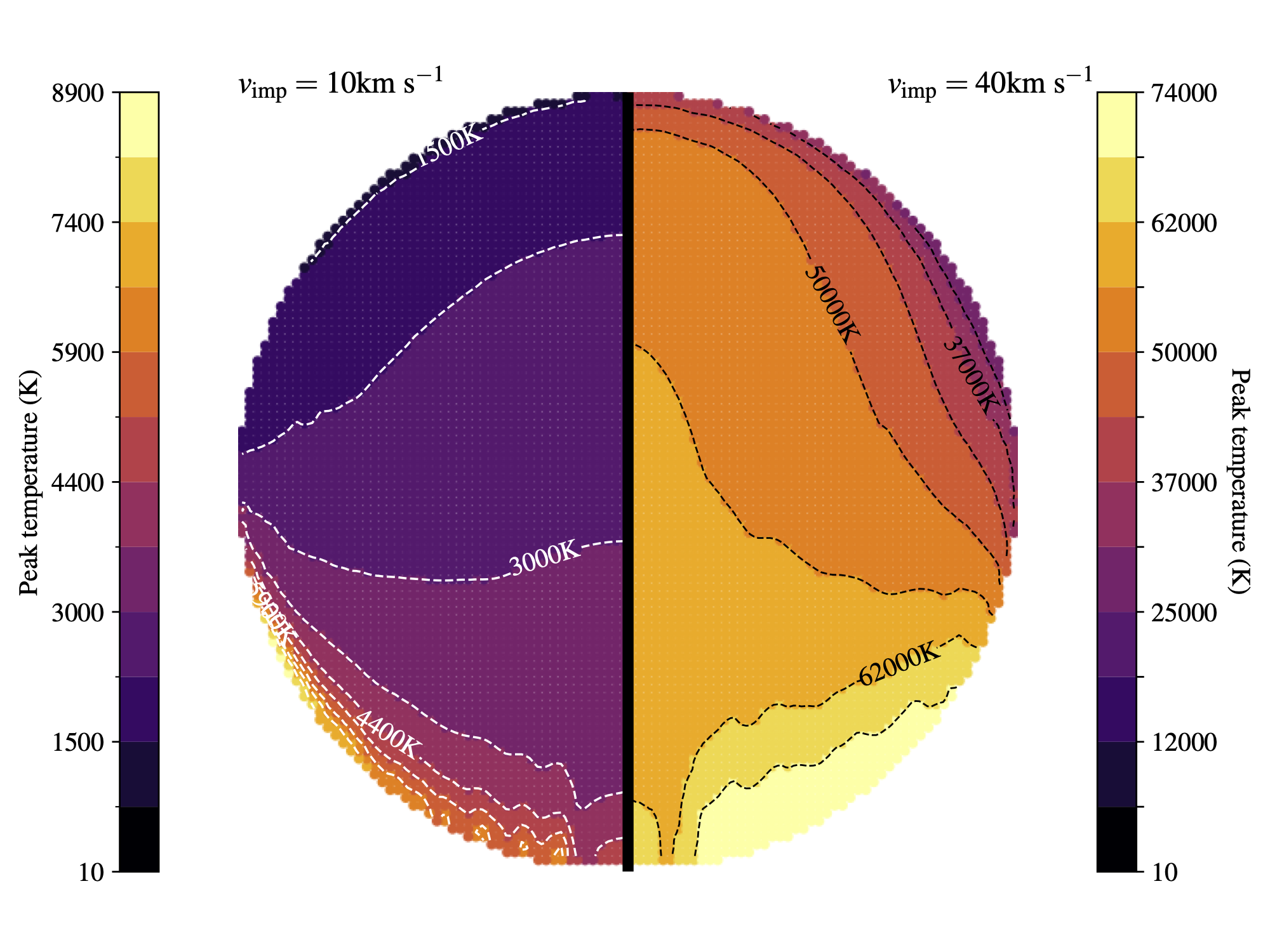}
    \caption{
    The peak temperature achieved by all tracer particles in simulations of a $1$~km diameter comet impacting at $10$~\kms (left) and $40$~\kms (right) in a provenance plot, where the tracers are shown at their initial locations prior to the impact. 
    A solid black line is shown between the two halves of the figure to delineate the difference between the two simulations. 
    The $10$~\kms impact on the left shows a peak temperature of $\sim 8900$~K, whereas the $40$~\kms impact on the right achieves a peak temperature of $\sim 74 \, 000$~K.
    Although the range of temperatures achieved during the two impacts is vastly different, the distribution of temperatures is similar, with a sharp increase in temperatures at the leading edge and raised temperatures towards the centre of the body. 
     }
    \label{fig:2D_velcompar_10_40}
\end{figure}
Figure~\ref{fig:2D_velcompar_10_40} shows the peak temperature recorded for each tracer particle, shown at their initial positions inside the comet, in simulations of a $10$~\kms impact (left) and a $40$~\kms impact (right). 
We use a discrete colour map to clearly show the temperature distribution within the comet and to be consistent with the results presented in figure~5 of \citetalias{PierazzoChyba1999:aminosurvival}.
The range of temperatures achieved during the two impact events are vastly different, at $v_\text{imp} = 10$~\kms only a very small region of the comet's leading edge exceeds $8900$~K, whereas almost the entire leading edge exceeds $\sim 74 \, 000$~K with the higher impact velocity.
However, the physical distribution of the temperature in the body is similar for both impact velocities.
In both cases, the highest temperatures are achieved at the leading edge and towards the centre of the body, with temperatures generally decreasing towards the top and outer edges of the body.
This is consistent with the temperature distribution shown in figure 5 of \citetalias{PierazzoChyba1999:aminosurvival}, although we note that the relatively reduced peak temperatures ($15 \, 000$~K) shown in their figure for a $20$~\kms impact is caused by their impact events occurring on an ocean layer and hence achieving lower shock wave amplitudes,.

\begin{figure}
    \centering
    \includegraphics[alt={The peak temperature experienced by an impacting body increases exponentially with impact velocity. There is a dark shaded region around the data which shows the interquartle range, which is very narrow and closely follows the data. A lighter shaded region shows the 5th and 95th percentile data, it also follows the same shape, but is much wider than the interquartile range. At 40 kilometres a second, the low temperature is approxiately 30000 Kelvin and the high temperature is around 80000 Kelvin, at 20 kilometres a second the low value is around 5000 Kelvin and the high temperature is around 22000 Kelvin.}, width=0.7\textwidth]{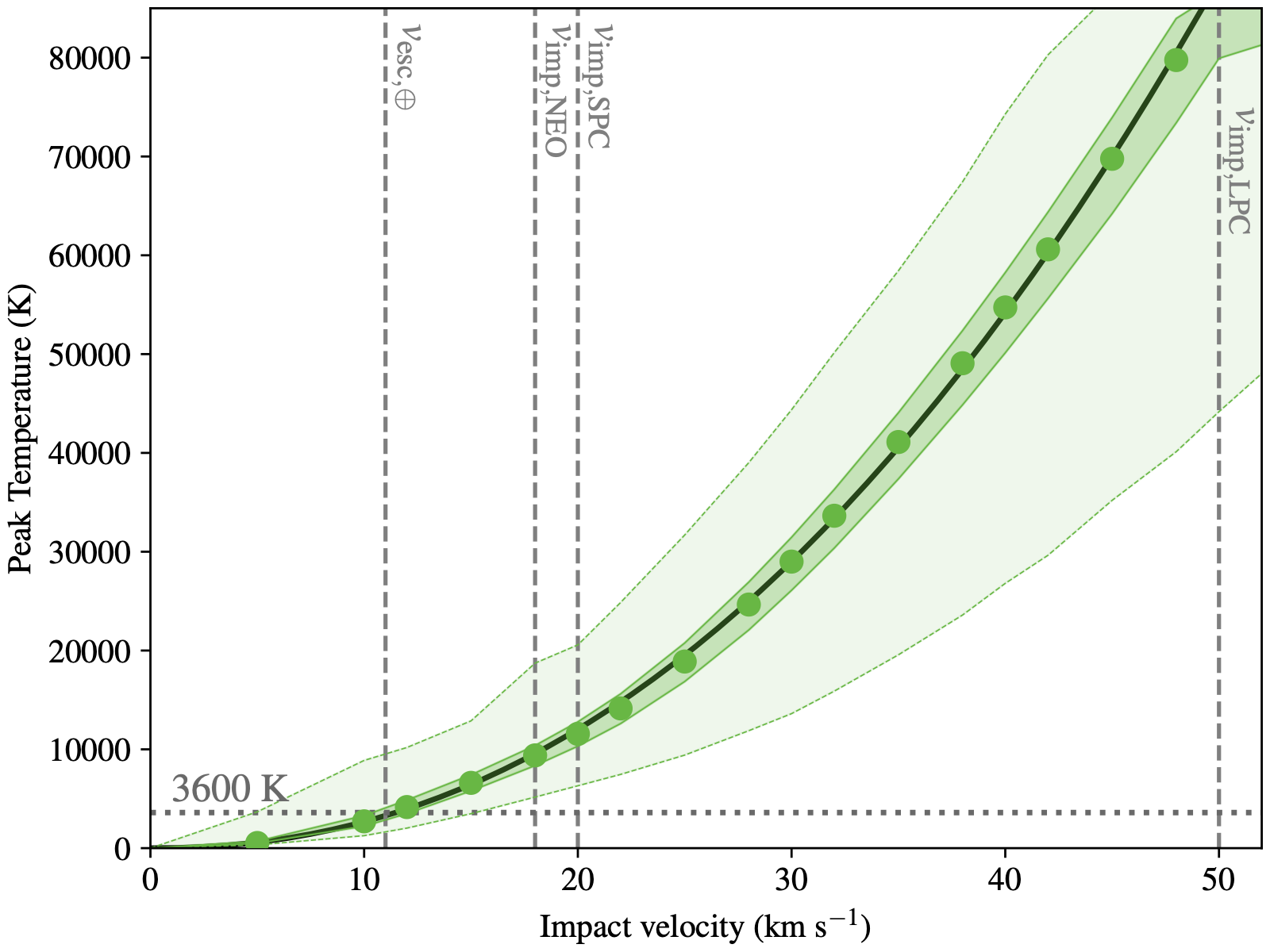}
    \caption{The median peak temperature experienced by all tracer particles inside a $1$~km diameter comet impacting with a range of impact velocities.
    The central shaded region bound by solid lines indicates the upper and lower quartiles of peak temperatures within the body, highlighting the narrow range of temperatures experienced by the bulk of the body. 
    The outer shaded region bound by the dashed lines indicates the 5th and 95th percentile of the peak temperatures showing the broad range of temperatures experienced by the rest of the projectile. 
    The vertical grey dashed lines show a range of velocities of particular interest, from left to right; the escape velocity of the Earth $v_{\text{esc,}\oplus} \sim 11$~\kms, the average impact velocity of near-Earth objects onto the Earth $v_{\text{imp, NEO}} = 18$~\kms \citep{Chesley2019:SyntheticImpactors}, the average impact velocity of short period comets $v_\text{imp, SPC}= 20$~\kms, and the average impact velocity of long period comets $v_\text{imp, LPC} = 50$~\kms \citep{HughesWilliams2000}.
    The solid black line shows a fit to the data as described in Equation~\ref{eq:Tmed_from_vimp}.
    The horizontal dotted line at $3600$~K, as discussed in Figure~\ref{fig:critHCNconc} is the approximate temperature at which OH dissociates and O becomes the major driver of HCN degradation.}
    \label{fig:Vel_temp}
\end{figure}
We further consider the effect of impact velocity on the distribution of peak temperatures recorded by the tracer particles within the impactor in Figure~\ref{fig:Vel_temp} which shows the median peak temperature and a range of peak temperature percentiles ($5$\textsuperscript{th}, $25$\textsuperscript{th}, $75$\textsuperscript{th}, $95$\textsuperscript{th}) as a function of impact velocity.
The majority of tracers experience temperatures close to the median with the interquartile range broadening at higher impact velocities. 
At lower impact velocities the remaining peak temperatures are skewed towards higher temperatures, whereas at higher impact velocities a broader range of low temperatures are recorded.
At an impact velocity of $45$~\kms, slightly less than the average impact velocity of a long-period comet ($\sim 50$~\kms), over half of the cometary material exceeds $\sim 70 \, 000$~K, an extreme temperature which would be coupled with a highly energetic, and damaging, impact event.
The horizontal, dotted, line at $3500$~K approximately represents the temperature from Figure~\ref{fig:critHCNconc} at which OH dissociates and 99\% of a sample of HCN will degrade within a fraction of a second, highlighting the difficulties of impacts with high velocities delivering HCN.
The solid black curve shown in Figure~\ref{fig:Vel_temp} shows a fit to the median peak temperature of the form $y = bx^c$, which allows us to parametrise the peak temperature $T_\text{med}$ (K) experienced during an impact as a function of impact velocity, $v_\text{imp}$ (\kms), as follows
\begin{equation}
    T_\text{med} = 18.02 \, v_\text{imp}^{2.17}.
    \label{eq:Tmed_from_vimp}
\end{equation}
An exponent of $\sim 2-3$ (as found here), is also expected from scaling laws for impact energy and melt production \citep[e.g.][]{AhrensOkeefe1977:Impacts, Pierazzo1997:MeltProduction, Manske2022:Melts}.
Although this expression does not encompass the full distribution of temperatures experienced by the impactor material, as the interquartile range closely and tightly follows the median, it is useful for considering the degree of heating as a function of impact velocity and the effect of impact angle which is further discussed in Section~\ref{subsec:impact_3D}.

We now consider the survival of HCN throughout the cometary body and its dependence on the impact velocity.
Following the discussion in Section~\ref{sec:chemistry}, we assume that the HCN is distributed homogeneously throughout the comet, with initial concentrations of \HCNconc$_0 \approx 5 \times 10^{19}$~\cmcubed and [\htwoo]$_0 \approx 3.3 \times 10^{22}$~\cmcubed in each grid cell occupied by cometary material.
At each simulation step, the temperature and pressure recorded by each tracer particle is used to calculate the instantaneous number density of OH in the tracer's grid cell assuming both steady state and equilibrium chemistries.
The instantaneous OH number density is then used in Equation~\ref{eq:HCN_degrate} with the temperature dependent reaction rate constant (Equation~\ref{eq:k_HCN}) to calculate the updated HCN concentration for the material within the tracer cell. 
\begin{figure}
    \centering
    \includegraphics[alt={The top panel shows a circle with 5 different markers pinpointing tracer particles spread throughout the body. The second panel shows the temperature of the tracer particles impulsively increasing shortly after the simulation started, the temperatures then slowly cool for the rest of the simulation. Tracers that are towards the lower edge of the body in the top panel reach, and retain, higher temperatures, all tracers keep elevated temperatures at the end of the simulation. The middle panel shows the pressure of the tracer particles increasing impulsively at the same time as the temperature increase. The pressures decay much quicker than the temperatures and they all return to their pre-impact pressure by 0.2 seconds. The final panel shows the percentage of HCN surviving the impact as a function of time. The tracers that are towards the bottom of the body lose all of their HCN with the impulsive increase in temperature and pressure. The tracer that is towards the rightside of the body about halfway up, loses about 60 percent of its HCN before the peak temperatures and pressures wane, and then retains this level of HCN. The tracer that is towards the trailing edge of the body and achieves the lowest temperatures and pressures, only loses a few percent of its HCN by the end of the simulation.}, width=0.48\textwidth]{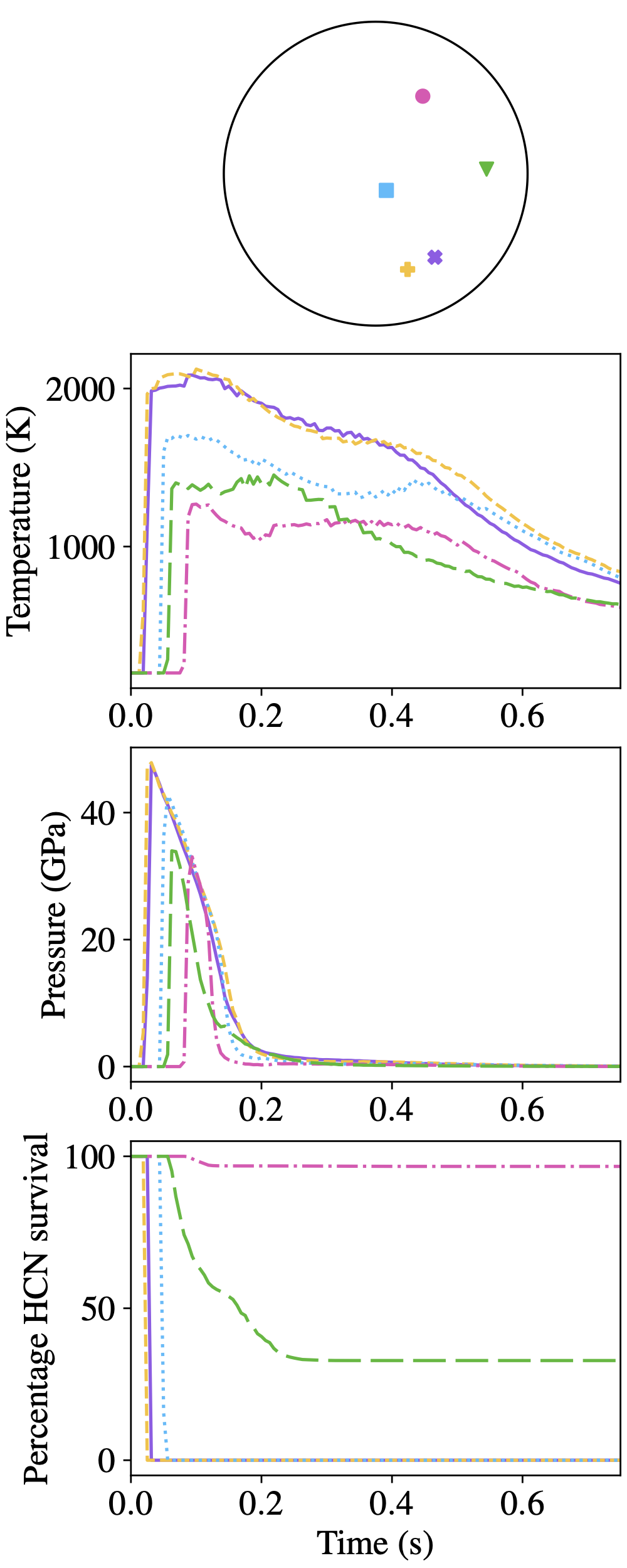}
    \caption{The initial location of five randomly selected tracer particles from a $1$~km diameter comet impacting vertically at $8$~\kms (top panel), the temperature profile of the tracer throughout the simulation (second panel), the pressure profile (third panel) and the surviving percentage of HCN within the tracer's material (bottom panel).
    The colour and line styles are consistent for each tracer through each panel.
    }
    \label{fig:TPSurvPaths}
\end{figure}
Figure~\ref{fig:TPSurvPaths} identifies the initial location within the cometary impactor of five randomly chosen tracer particles, the temperatures and pressures experienced by the tracers throughout the simulation, and the corresponding surviving percentage of HCN. 
Tracers towards the leading edge and towards the centre of the projectile experience the largest pressures and temperatures (as in Figure~\ref{fig:2D_velcompar_10_40}), and hence the largest and quickest degradation of HCN. 
Two tracer particles towards the top and right hand side of the projectile, experience partial HCN degradation while the tracer material remains in a shocked state. 
Once the tracers have been released from the shock the destruction ceases, indicating that the initial shock pulse is the most disruptive for HCN survival in the initial impact stage.

\begin{figure}
    \centering
    \includegraphics[alt={The left side of this figure is the same as figure 3, where the highest temperatures of around 7000 Kelvin are felt at the leading edge of the body, and the temperatures increase to around 1500 Kelvin towards the trailing edge. The right hand side of the figure indicates that there is essentially 0 survival of HCN throughout the majority of the body, with a very thin region of high survival above 60 percent at the trailing edge which aligns with the region in the left panel achieving 1500 Kelvin.}, width=0.7\textwidth]{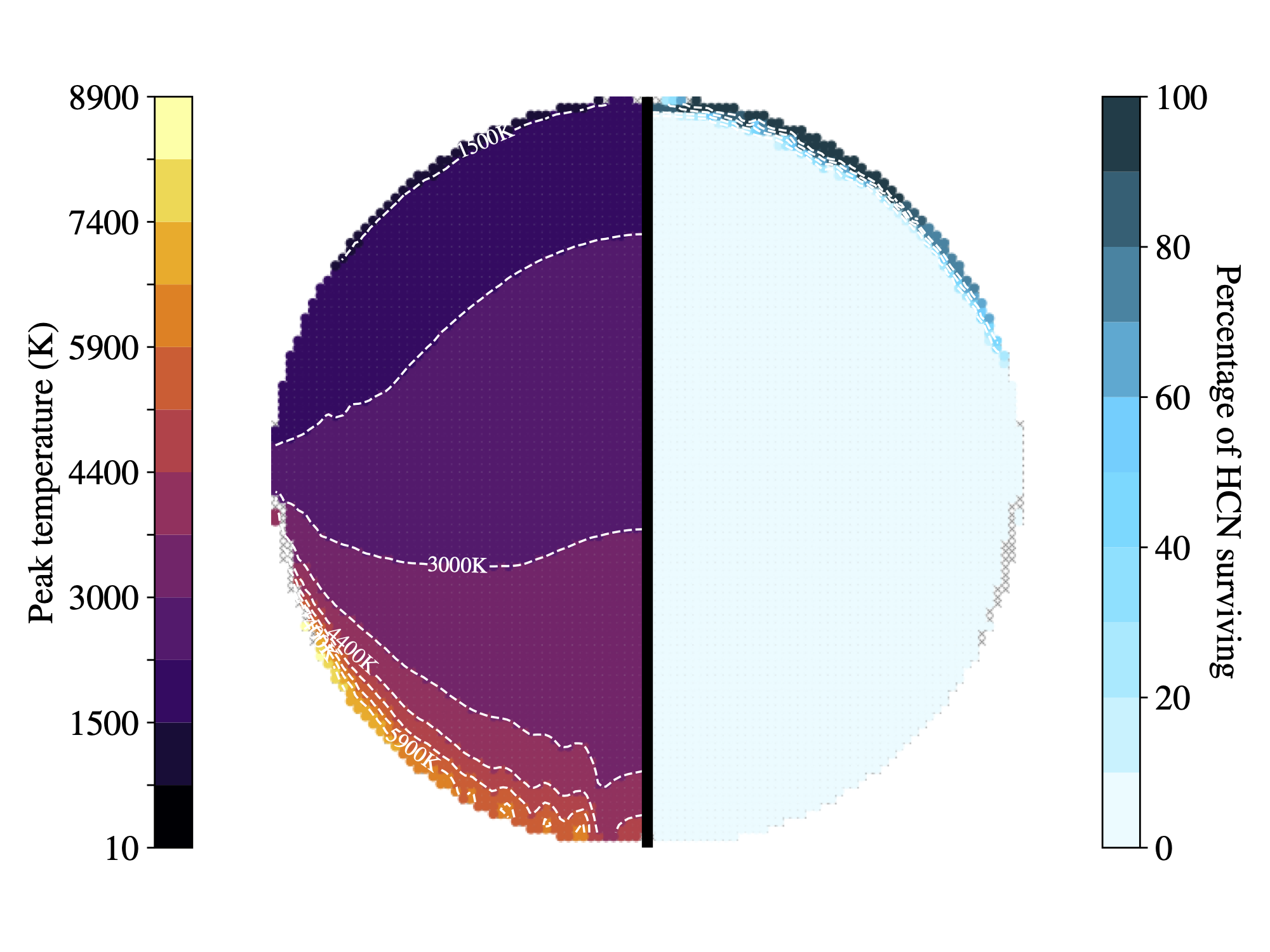}
    \caption{The peak temperature achieved by all tracer particles within a $1$~km diameter comet during a $10$~\kms impact in a provenance plot (where the tracers have been placed back in their original locations within the impactor), the peak temperatures are described by the left colour bar.
    The right-hand side of the figure shows the percentage of HCN surviving (assuming equilibrium concentrations of OH) given the temperatures and pressures recorded by the tracer particles during the course of the simulation. 
    The small grey crosses visible in the right-hand side of the plot indicate the initial locations of tracers which are removed during the course of the simulation due to drifting out of the grid cell containing the tracked cometary material (see Section~\ref{subsec:impact_2D}).
    At this impact velocity, HCN survival is only possible at the trailing edge of the impactor which achieves relatively low temperatures for a very short period of time.
    }
    \label{fig:HCN_spatial_surv}
\end{figure}
Figure~\ref{fig:HCN_spatial_surv} displays the survival of HCN throughout a $1$~km diameter comet impacting at $v_\text{imp} = 10$~\kms, assuming \OHconc is in equilibrium.
As in Figure~\ref{fig:2D_velcompar_10_40} the data for each tracer particle is displayed at the particle's initial position within the impactor. 
The left hand side of the figure shows the peak temperature recorded by the tracer particle across the duration of the simulation and the right hand side of this figure shows the percentage of HCN surviving at the end of the simulations.
Here, we define the surviving percentage as $[\text{HCN}]_f / [\text{HCN}]_0 \times 100$, where the $f$ subscript denotes the final HCN concentration.
It can be clearly seen that the survival of HCN closely traces the temperature distribution of the material, with the survival of HCN only being possible at the lowest peak temperatures at the trailing edge of the projectile.

\begin{figure}
    \centering
    \includegraphics[alt={The percentage of HCN surviving an impact as a function of impact velocity increases very rapidly from 0 to 100 percent between 15 and 8 kilometres a second for the kinetic chemistry model. Survival increases from 0 to 100 percent between 11 and 5 kilometres a second for the equilibrium model.}, width=0.7\textwidth]{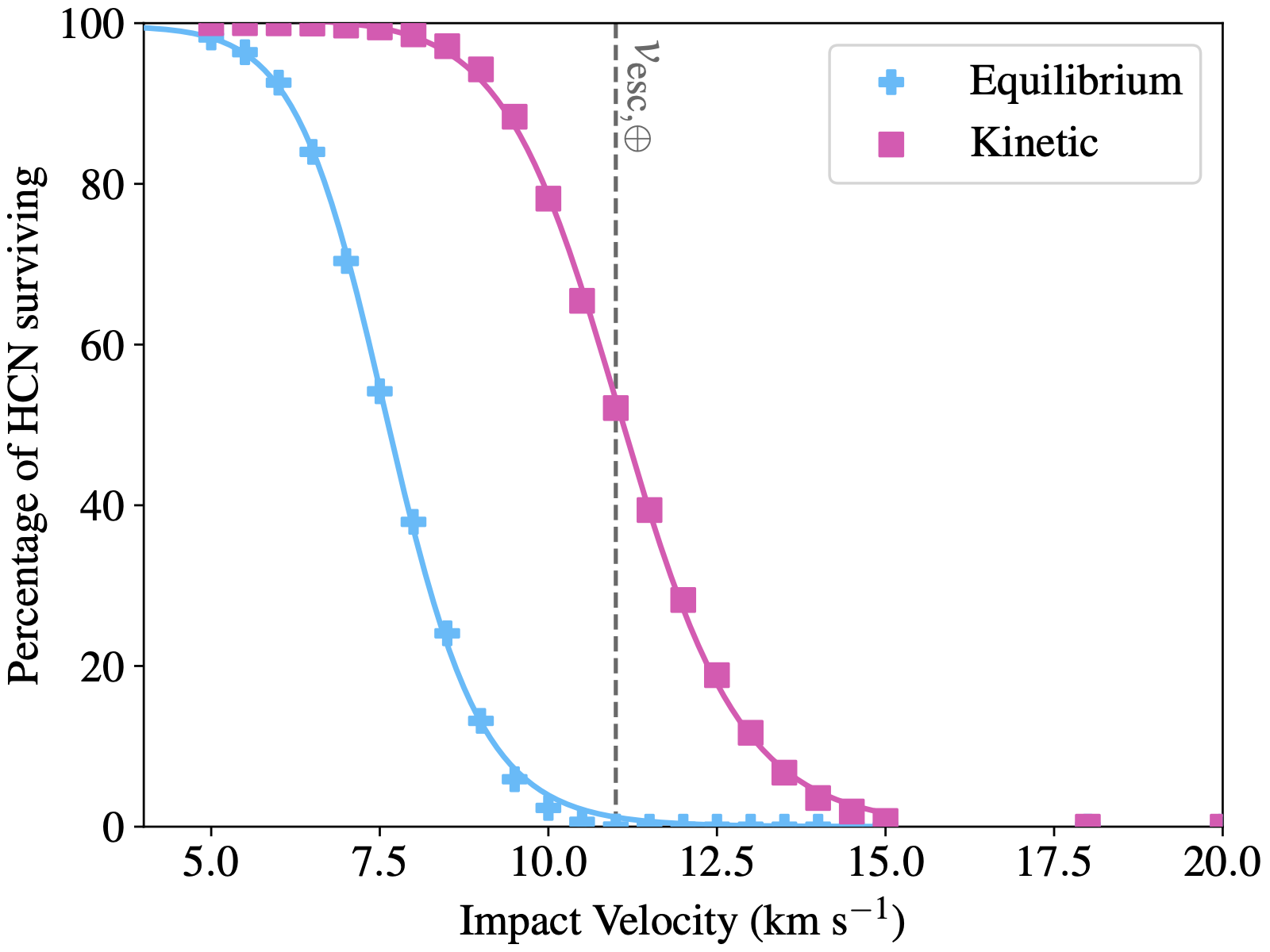}
    \caption{The total percentage of HCN inside a $1$~km diameter comet which survives a vertical impact as a function of impact velocity. 
    In blue crosses we show the survival assuming equilibrium chemistry and in pink squares we show the survival assuming steady state kinetic chemistry.
    The vertical dashed grey line shows the approximate escape velocity of the Earth $v_{\text{esc, } \oplus} \sim 11$~\kms, indicating the difficulties in successfully delivering HCN in impacts, especially \OHconc is in equilibrium.
    The coloured solid lines show a fit to the survival data in the form of a slightly modified logistic function as described in Equation~\ref{eq:surv_vimp}.
    }
    \label{fig:vel_HCNsurv}
\end{figure}
Figure~\ref{fig:vel_HCNsurv} takes a closer look at the effect of impact velocity on the prospects of HCN survival by examining the total HCN survival throughout a $1$~km diameter comet as a function of impact velocity, assuming both equilibrium and kinetic steady state chemistry.
It is clear that for either chemistry, any impact with $v_\text{imp} > 15$~\kms will see zero HCN survival, for equilibrium chemistry the prospects are even worse with survival only being possible for $v_\text{imp} < 11$~\kms.
As in Section~\ref{sec:chemistry}, it is likely that the true survival rate of HCN as a function of impact velocity lies closer to the equilibrium chemistry curve than kinetic steady state and that our models overestimate survival for both chemistries.
Given that $v_{\text{esc, }\oplus} \sim 11$~\kms is the minimum unaltered impact velocity that an impactor on the Earth could achieve, we note that impacts below $10$~\kms require strong aerobraking in the atmosphere prior to impact, and we do not take into account any effect this may have on the comet chemistry prior to impact.

The survival of HCN as a function of impact velocity, $v_\text{imp}$ (\kms), can be described by a slightly modified logistic function as follows 
\begin{equation}
    \gamma (v_\text{imp}) = \frac{A}{1 + \text{exp} \left\{ q \left( v_\text{imp}^{1/2} - v_0^{1/2} \right) \right\}},
    \label{eq:surv_vimp}
\end{equation} 
where $\gamma (v_\text{imp})$ is the surviving percentage of HCN, $A = 99.64 \, (100.22)$, $q = 7.99 \, (7.53)$~s$^{1/2}$~km$^{-1/2}$ and $v_0 = 7.63 \, (11.10)$~km~s$^{-1}$ are parameters determined by fitting the survival for equilibrium (and kinetic) chemistry models. 

Thus, survival of HCN above $1$~\% in a $1$~km diameter comet is only possible for $v_\text{imp} < 15$~\kms assuming steady state kinetic chemistry and $v_\text{imp} < 11$~\kms assuming equilibrium  chemistry, both velocity limits are close to the escape velocity of the Earth.
Below these velocities, the survival of HCN increases rapidly with decreasing impact velocity.

\subsubsection{Comet Size} \label{ssubsec:2D_size}

Now we turn again to observations of the Solar System to investigate the effect of the impactor's size on the survival of HCN.
Using the Pan-STARRS1 near-Earth object survey, \cite{Boe2019:LPC_SFD} identify a size-frequency distribution (SFD) for the Solar System's long period comets (LPCs) which shows a relatively steep slope for LPCs $\gtrsim 3$~km, and a shallower slope for smaller objects. 
Taking into account the detection efficiency of the Pan-STARRS survey, \cite{Boe2019:LPC_SFD} estimate that there are $\sim 0.46 \times 10^{9}$ active LPCs larger than $1$~km with a perihelion below $10$~au.

For smaller sized impactors, we again utilise the atmospheric entry model as described in \cite{Anslow2025:atmosentry}, which finds comets with a diameter of $\sim 300$~m at the top of the atmosphere, lose some mass during passage through the atmosphere but are able to survive and impact the surface of the Earth.
Thus, we consider comets with $0.5$~km $< d < 10$~km as being able to impact the surface of the planet without large scale atmospheric processing or excessive damage to the local environment \citep[e.g][]{MaherStevenson1988:ImpactFrustration, AbramovMojzsis2009:HadeanHab, Abramov2013:HadeanImpacts} and investigate their impacts through simulations.

At the point of impact, shock waves are simultaneously sent down into the impacted surface and upwards throughout the body of the impactor, raising the temperature and pressure of the material as it passes. 
The impactor material is released from its shocked state by the propagation of rarefaction waves from the top of the projectile. 
Thus, the time the material stays in its shocked state with heightened temperatures depends on the size and velocity of the impactor \citep[e.g.][]{Melosh1989:ImpactCratering}.
The scaled time of our simulations captures this shock and release process and scales linearly with impactor diameter.
If only the size of the impactor changed, the distribution of temperatures within the projectile would remain constant, but the duration of the shocked state would be shorter for smaller impactors.

We carry out simulations of impactors with diameters in the range $0.5$~km $< d < 10$~km with fixed CPPR = 50 at a smaller range of impact velocities than considered in Section~\ref{sssec:ImpactVelocity}.
\begin{figure}
    \centering  
    \includegraphics[alt={Normalised to the survival of HCN in a 1 kilometre diameter body, the survival as a function of impactor diameter is constant for a 5 kilometre a second impact. In a 8 kilometre a second impact the survival decreases shallowly with impactor diameter. Finally, in a 10 kilometre a second impact, survival decreases more steeply with increasing diameter, with a 10 kilometre diameter comet seeing only 40 percent of the HCN surviving an impact of a 1 kilometre diameter comet.}, width=0.7\textwidth]{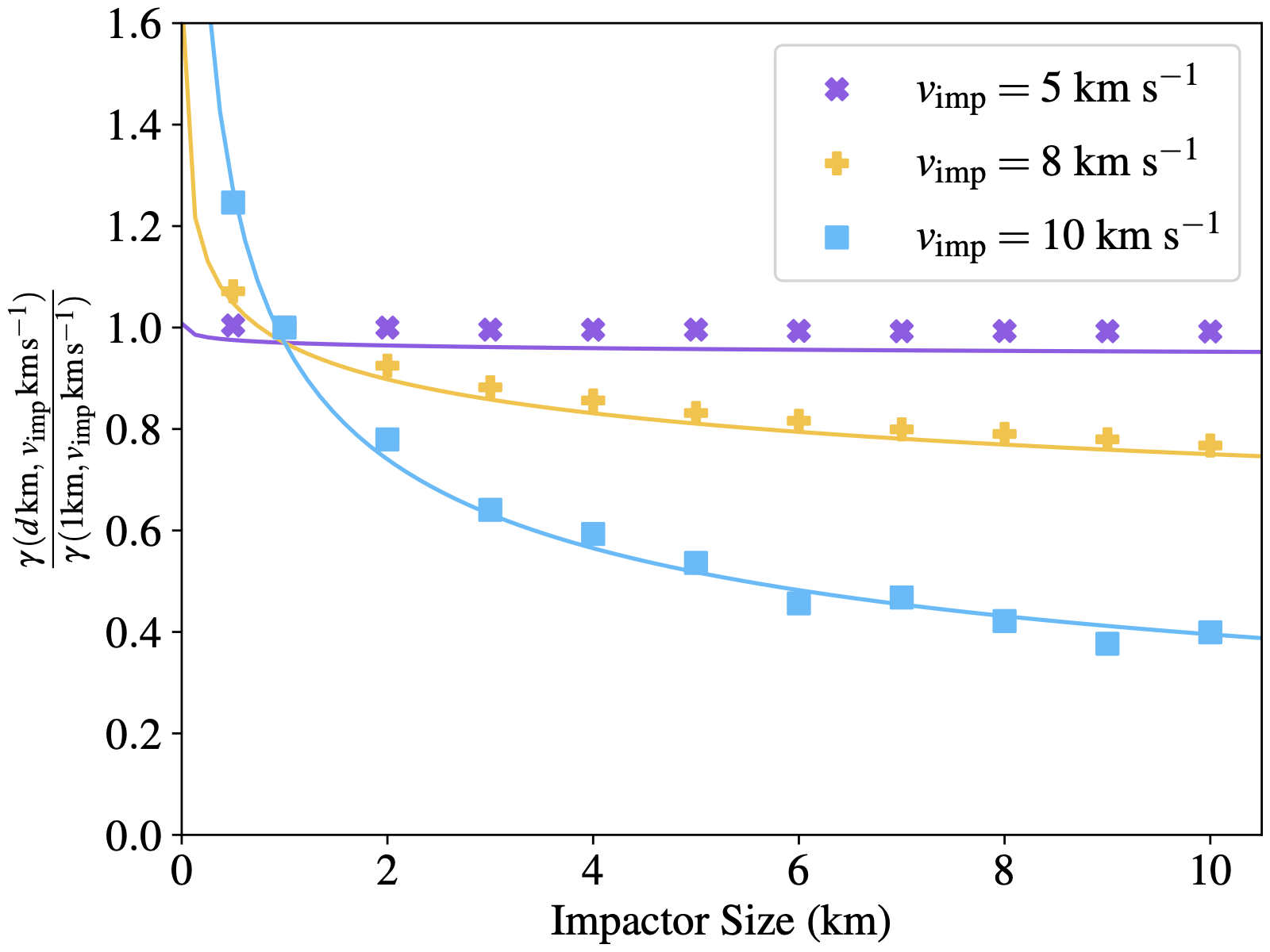}
    \caption{The percentage of HCN which survives a cometary impact at a range of impact velocities (as described in the legend) assuming equilibrium chemistry as a function of impactor diameter normalised to the percentage survival for a $d = 1$~km cometary impactor. 
    The percentage survival for a $1$~km diameter comet for the different velocities are as follows: $\gamma \, (1\text{km}, \, 5\text{km s}^{-1}) = 97.86\%$, $\gamma \,(1\text{km}, \, 8\text{km s}^{-1}) = 37.81\%$, and $\gamma \, (1\text{km}, \, 10\text{km s}^{-1}) = 2.21\%$.
    The solid lines shows a fit to the data of the form $\gamma\, (d, v) / \gamma \, (1\text{km}, v) \sim 0.97 d^{\Gamma\, (v)}$, where $\Gamma \, (v) = (0.085 v_\text{imp})^{5.61}$.}
    \label{fig:SizeSurvival}
\end{figure}
Figure~\ref{fig:SizeSurvival} shows the percentage of HCN which survives an impact of various velocities as a function of impactor diameter, normalised to the percentage survival for a $1$~km diameter comet. 
We chose to normalise the data to a $1$~km diameter impactor in order to align with the fiducial impactor size used throughout the rest of our simulations.
As expected, the survival of HCN decreases with increasing diameter and increasing time spent in a shocked state, with higher impact velocities showing a stronger effect.
For all impact velocities the survival rate plateaus, indicating that all HCN which will degrade during the impact has decayed by the time the shock wave has passed through the body (as in Figure~\ref{fig:TPSurvPaths}) and increased temperatures (higher impact velocities) would be required to further reduce the survival. 

The survival of HCN normalised to a $1$~km diameter impactor can be parametrised as a function of impactor diameter and velocity as follows,
\begin{equation}
    \frac{\gamma \left( d \, \text{km}, \, v_\text{imp} \, \text{km s}^{-1}\right)}{\gamma \left(1 \, \text{km}, \, v_\text{imp}  \, \text{km s}^{-1}\right)} = (B \, d)^{\Gamma(v_\text{imp})},
    \label{eq:size_scaling}
\end{equation}
where,
\begin{equation}
    \Gamma (v_\text{imp}) = - (C v_\text{imp})^{E}.
    \label{eq:sizescaling_velparam}
\end{equation}
The fitting parameters determined from the simulation data are $B = 0.97 \, (0.99)$~km$^{-1}$, $C = 0.085 \, (0.068)$~s km$^{-1}$, and $E = 5.61 \, (6.63)$ for equilibrium (kinetic steady state) chemistry.
This relationship is shown in the solid lines in Figure~\ref{fig:SizeSurvival} for equilibrium chemistry,.
Thus, the surviving percentage of HCN in a comet of any diameter impacting vertically can be estimated by finding the percentage survival for a $1$~km object at the required impact velocity as shown in Figure~\ref{fig:vel_HCNsurv} and then evaluating Equations~\ref{eq:size_scaling}-\ref{eq:sizescaling_velparam}.

\subsection{3D Simulations} \label{subsec:impact_3D}
The final parameter we consider in our assessment of the survivability of HCN in cometary impacts is the impact angle $\theta \, (^\circ)$.
Although all of the simulations discussed previously consider vertical impacts, the most likely angle of impact is $45^\circ$, $50$~\% of impacts will occur within the range $30^\circ - 60^\circ$ and $76.7$~\% will occur with $20^\circ < \theta < 70^\circ$ \citep{Gilbert1893:LunarFace, Shoemaker1962, PierazzoMelosh2000a:ObliqueReview}. 
Thus, the vertical $90^\circ$ impacts considered in Section~\ref{subsec:impact_2D} are unlikely, and the obliquity of the impact (angle from the horizontal) must be taken into account.
Given that oblique impacts produce asymmetric shock waves and ejecta distributions, they can no longer be modelled by two dimensional simulations, which introduces an additional complication to computationally modelling realistic cometary impacts. 

\cite{PierazzoMelosh2000b:ObliqueHydro} carried out a series of two- and three-dimensional simulations to model the Chicxulub impact event.
They find that  two-dimensional vertical simulations underestimate the peak shock pressure, but overestimate the peak temperatures, implying that the HCN survival rates calculated in Section~\ref{subsec:impact_2D} are overestimates.
The results of \cite{PierazzoMelosh2000b:ObliqueHydro} suggest that the peak shock temperatures experienced during a simulation scale with impact angle with a factor $\text{sin} \, \theta \, ^{3/2}$ and the post-shock temperature scales with $\text{sin} \, \theta \, ^{0.8}$. 
\citetalias{PierazzoChyba1999:aminosurvival} use these scaling laws to investigate the effect of impact angle on the survival of amino acids in their work, multiplying all of their vertical impact tracer temperatures by $\text{sin} \, \theta \, ^{0.8}$, similarly \citetalias{ToddOberg2020:HCNcomet} uses this scaling to calculate the survival of HCN.

Exploiting two decades of advances in computational capabilities, we produce new three-dimensional simulations to explore the effect of impact obliquity on the survival of HCN in cometary impacts and validate the temperature scaling laws introduced in \cite{PierazzoMelosh2000b:ObliqueHydro} and used in \citetalias{PierazzoChyba1999:aminosurvival} and \citetalias{ToddOberg2020:HCNcomet}.
We can exploit the symmetry of our spherical comet by only simulating half of the projectile. 
Utilising the same CPPR of 50 and placing a tracer particle in each grid cell which includes projectile material, increases the number of tracers from 3923 in our 2D simulations to 259172.
For our 3D simulations we explore a smaller parameter space informed by our wider exploration in Section~\ref{subsec:impact_2D}, sampling the following impact angles $\theta = [15, \, 30, \, 45, \, 60, \, 75, \, 90]^\circ$ at a restricted range of impact velocities, $v_\text{imp} = [8, \, 10, \, 12, \, 15, \, 20, \, 30]$~\kms all for a $1$~km diameter impactor.

\subsubsection{Impact Angle}

Equipped with new higher resolution tracer temperature and pressure histories, we are able to reassess the temperature scaling laws presented in \cite{PierazzoMelosh2000b:ObliqueHydro} and used in \citetalias{PierazzoChyba1999:aminosurvival} and \citetalias{ToddOberg2020:HCNcomet}.
\begin{figure*}[t!]
    \centering
    \begin{subfigure}[t]{0.45\textwidth}
        \caption{}
        \includegraphics[alt={The peak temperature experienced during a comet impact decreases as a function of impact angle, for all impact velocities. A shaded region around the data point shows the interquartile range which follows the same decreasing pattern, is broadest at angles between 60 degrees and 30 degrees and narrows at 90 degrees and 15 degrees. The fitting line for sine theta squared underestimates temperatures for all impact angles. The fitted line for sine theta to the power of 1.6 follows the same distribution of the data and lies approximately between the data for all of the impact velocities. }, width=0.88\textwidth]{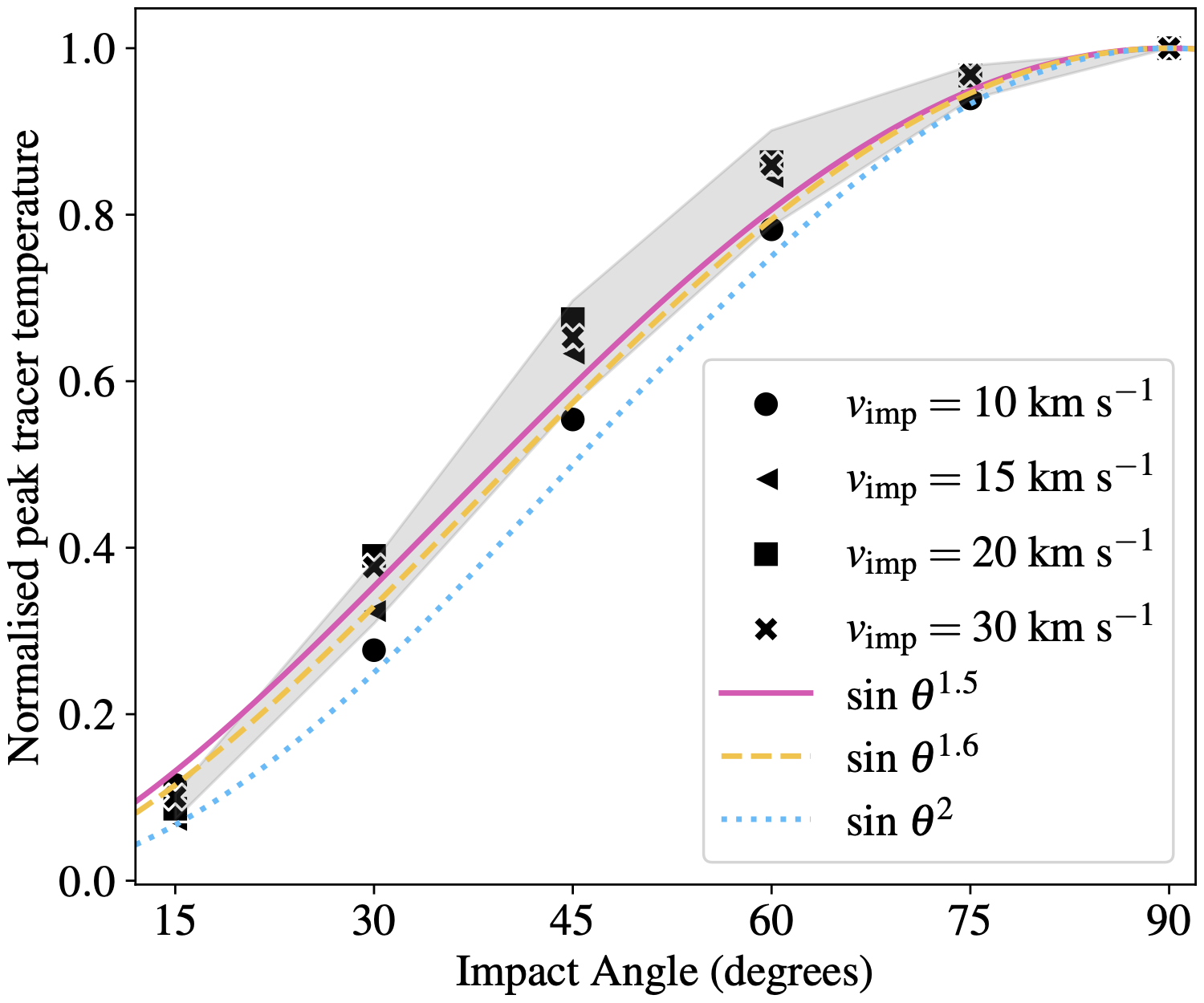}
        \label{subfig:normpeakT}
    \end{subfigure}
    \begin{subfigure}[t]{0.455\textwidth}
        \caption{}
        \includegraphics[alt={The temperature recorded at the end of our simulations decreases with decreasing impact angle. Lower impact velocities have lower post-impact temperatures. The interquartile range is broadest around impact angles between 60 and 30 degrees, and narrows at 15 degrees. The previously used relationship of sin theta to the power of 0.8 only describes the data for a 30 kilometre a second impact above 30 degrees. The new expression we describe in Equation 13 provides a better fit to the post-impact temperatures for all impact velocities.}, width=0.9\textwidth]{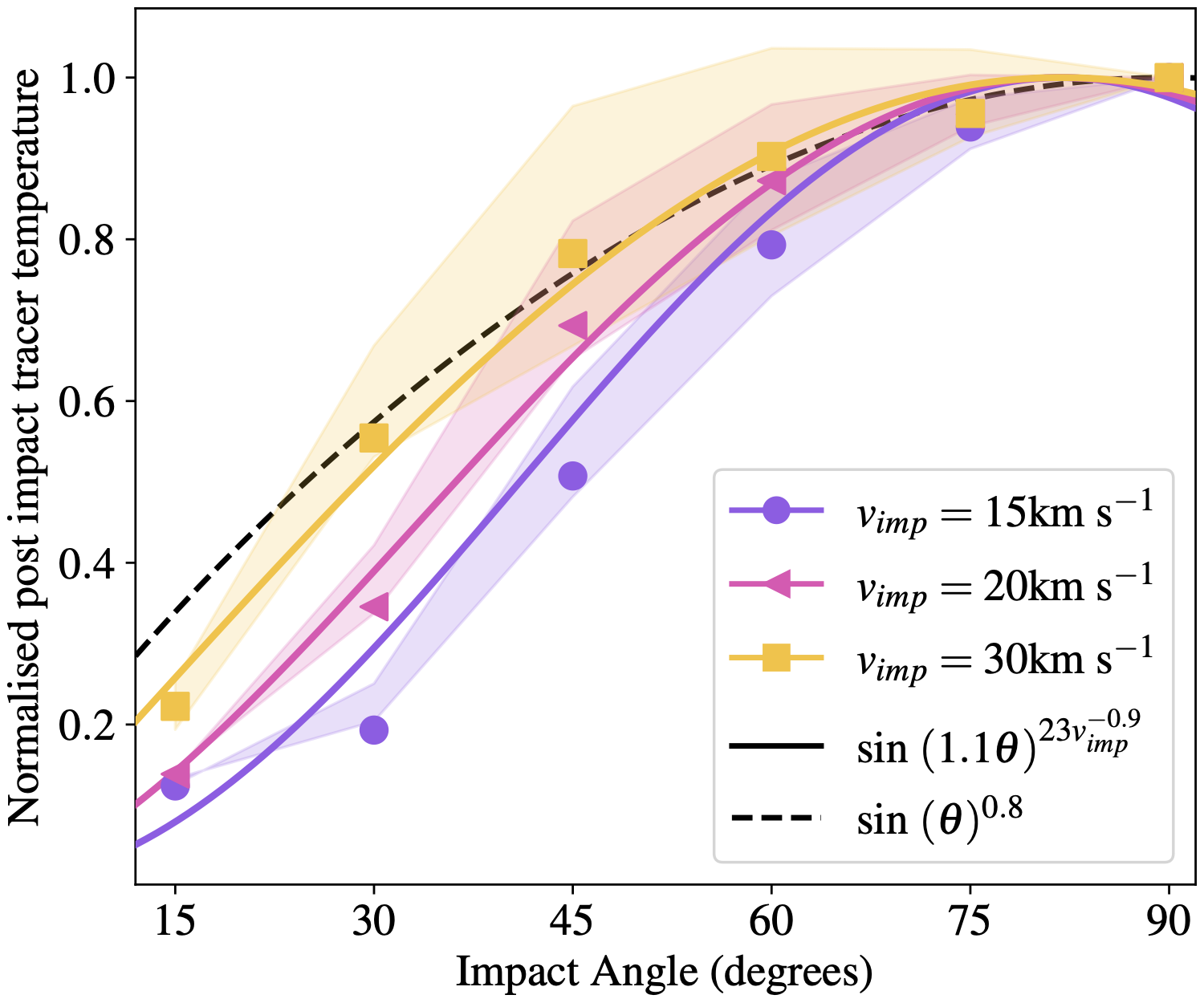}
        \label{subfig:normpostT}
        \end{subfigure}
    \caption{(a) The peak temperatures recorded by tracer particles in simulations for a range of impact angles, normalised by the mean peak tracer temperature recorded in a vertical ($\theta = 90^\circ$) impact. 
    The markers indicate the median peak temperature recorded in simulations with impact velocity as shown in the legend.
    The grey shaded region shows the normalised combined interquartile ranges for all impact velocities.
    The coloured curves indicate various scaling relations for the impact angle which have been used previously as described in the legend and further in the text.
    (b) The post-impact tracer temperature for a range of impact angles normalised to the post-impact tracer temperature for a vertical impact.
    Again, the markers show the median post-impact temperature and and the shaded region shows the normalised interquartile range for temperatures.
    The previously used scaling relation for the post-impact temperature ($\text{sin} \, \theta\, ^{0.8}$) is shown by the dashed black line.}
    \label{fig:angleScaling}
\end{figure*}

Figure~\ref{subfig:normpeakT} shows the median peak temperature recorded across all tracers in a $1$~km diameter comet, normalised to the temperature of a direct vertical impact ($\theta = 90^\circ$), as a function of impact angle, for a range of impact velocities. 
The previously used scaling relation is shown in the solid pink line (e.g. \citealt{PierazzoMelosh2000b:ObliqueHydro}, \citetalias{PierazzoChyba1999:aminosurvival} and \citetalias{ToddOberg2020:HCNcomet}).
Although on average the peak temperatures experienced during impacts can be reasonably well described by the previously used relation ($\text{sin} \,  \theta \, ^{1.5}$), here we suggest a slightly altered scaling relation as follows 
\begin{equation}
    T_\text{peak}(\theta) = T_\text{peak}(90^\circ) \, \text{sin} \, \theta \, ^{1.6},
    \label{eq:Tmean_theta_scale}
\end{equation}
which we find better encompasses low impact velocities and angles.
The grey shaded region in Figure~\ref{subfig:normpeakT} shows the normalised interquartile range for all impact velocities shown in the figure which narrows with lower impact angle, indicating that our scaling relations are able to encompass the variation of temperatures within the impactors as well as they are able to encompass the differences caused by impact velocities.

Similarly, Figure~\ref{subfig:normpostT} shows the median post-impact tracer temperature at the end of the simulation time, normalised to a vertical impact, for a range of impact velocities and angles. 
The shaded regions again show the interquartile range of post-impact temperatures experienced by the different impact velocities, which narrows with lower impact angle.
The previously used scaling relation for the post-impact temperature $T_\text{post}(\theta)~=~T_\text{post}(90^\circ)~\text{sin}~\theta~^{0.8}$ (which is shown by the dashed black line in Figure~\ref{subfig:normpostT}; \citealt{PierazzoMelosh2000b:ObliqueHydro}), only appears to resemble the angle scaling for the highest impact velocity $30$~\kms down to a $\theta = 30^\circ$ impact.
At the lowest impact angle we consider ($15^\circ$), the previous post-impact temperature scaling law overestimates temperatures by around $10$~\%.
For lower impact velocities (e.g. $15$~\kms) and angles (e.g. $30^\circ$) this scaling law overestimates post-shock temperatures by around $40$~\%.
We find that the post-impact temperature is sensitively dependent on the impact velocity, and suggest a scaling relation as follows to better represent this dependency
\begin{equation}
    T_\text{post}(\theta) = T_\text{post}(90^\circ) \, \text{sin} \, (1.1 \theta) \, ^{23 v_\text{imp}^{-0.9}}.
    \label{eq:Tmeean_post_theta_scale}
\end{equation}
However, we note that the post-impact temperature is entirely dependent on how `post-impact' is defined.
The post-shock temperatures reported in \citetalias{PierazzoChyba1999:aminosurvival} are taken $4-4.5$~s after the impact, beyond the scaled time of our high impact velocity and small diameter simulations.
Results presented in \cite{Halim2021:MoonBiomarkers} for impact simulations of porous projectiles impacting solid targets show large differences between peak and post-impact temperatures in their figure 5.
However, their post-impact temperatures are recorded $5$~ms after impact, again highlighting the study-dependent definition of post-impact temperature.
Further, the previous work estimating the survival of HCN (or amino acids) in cometary impacts presented in \citetalias{PierazzoChyba1999:aminosurvival} and \citetalias{ToddOberg2020:HCNcomet} uses the scaling factor $(\text{sin} \, \theta)^{0.8}$ (as for post-impact temperatures above) to adjust their temperature histories to account for impact obliquity, although \citetalias{PierazzoChyba1999:aminosurvival} acknowledged this as a conservative correction as the results of \cite{PierazzoMelosh2000b:ObliqueHydro} suggests this overestimates peak shock temperatures.

Therefore, we chose to move away from using the ambiguous post-shock temperatures and focus on the more accurate relationship between peak temperatures and impact angle as shown in Figure~\ref{subfig:normpeakT}.
\begin{figure*}[t!]
    \centering
    \begin{subfigure}[t]{0.45\textwidth}
        \caption{}
        \includegraphics[alt={The survival of HCN using the equilibrium chemical model follows an s-shaped distribution as a function of impact angle for all impact velocities, with lowest survival at higher impact angles, rapidly increasing survival at intermediate angles, and survival plateauing towards 100 per cent at the lowest simulated impact angles of 15 degrees. For each impact velocity, a dashed line also shows the results of equations 14 and 15, which does a good job of following the data for a 10 and 12 kilometre a second impact. At the slowest impact velocity of 8 kilometres a second, the parametrisation underestimates survival at the highest impact angles, and slightly overestimates at the lower impact angles.}, width=0.88\textwidth]{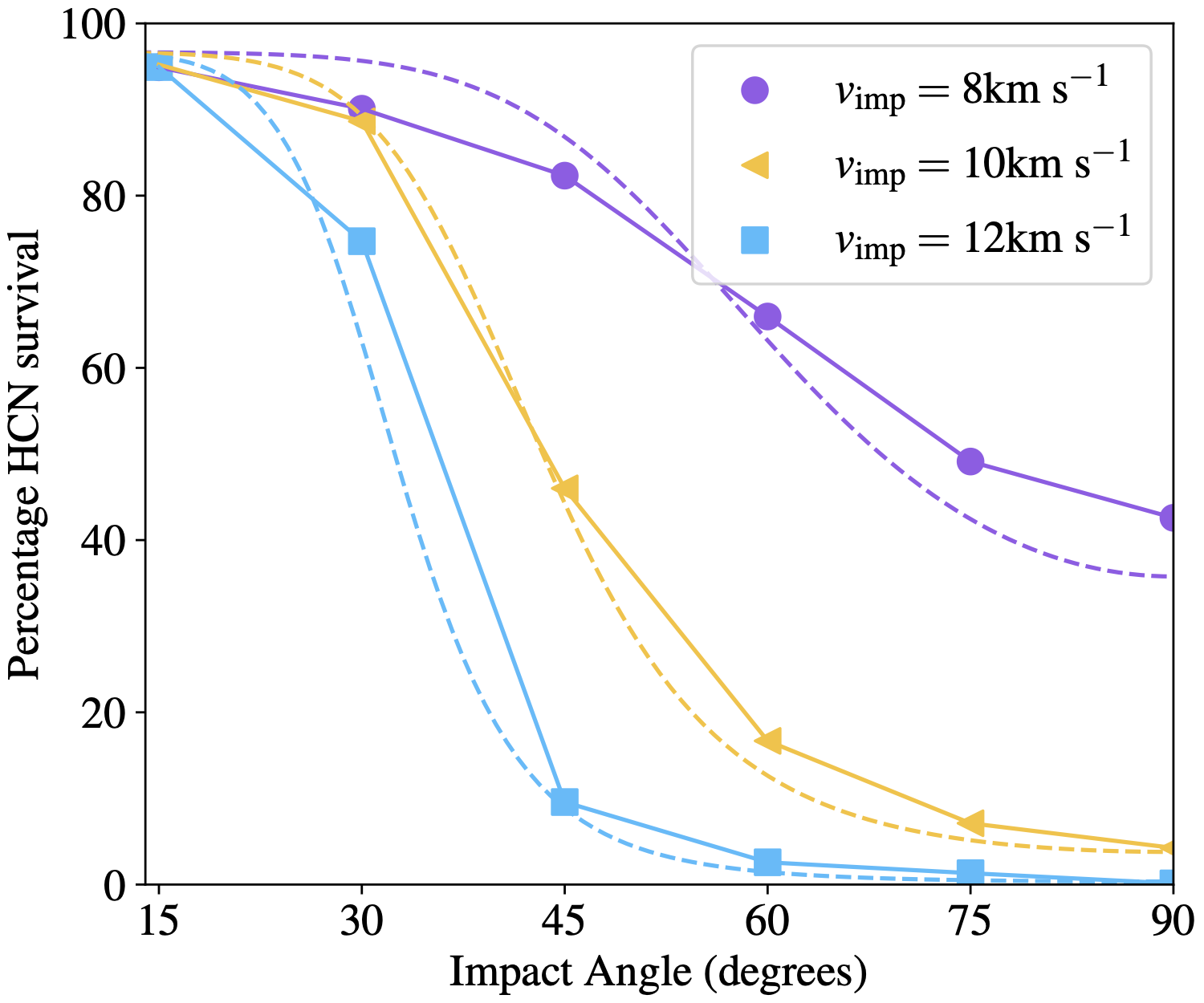}
        \label{subfig:full_equil}
    \end{subfigure}
    \begin{subfigure}[t]{0.455\textwidth}
        \caption{}
        \includegraphics[alt={This figure is the same as the left panel but for the kinetic chemistry, the data follows the same shape for all impact velocities. This panel considers the impact velocities 12, 15 and 20 kilometres a second.}, width=0.9\textwidth]{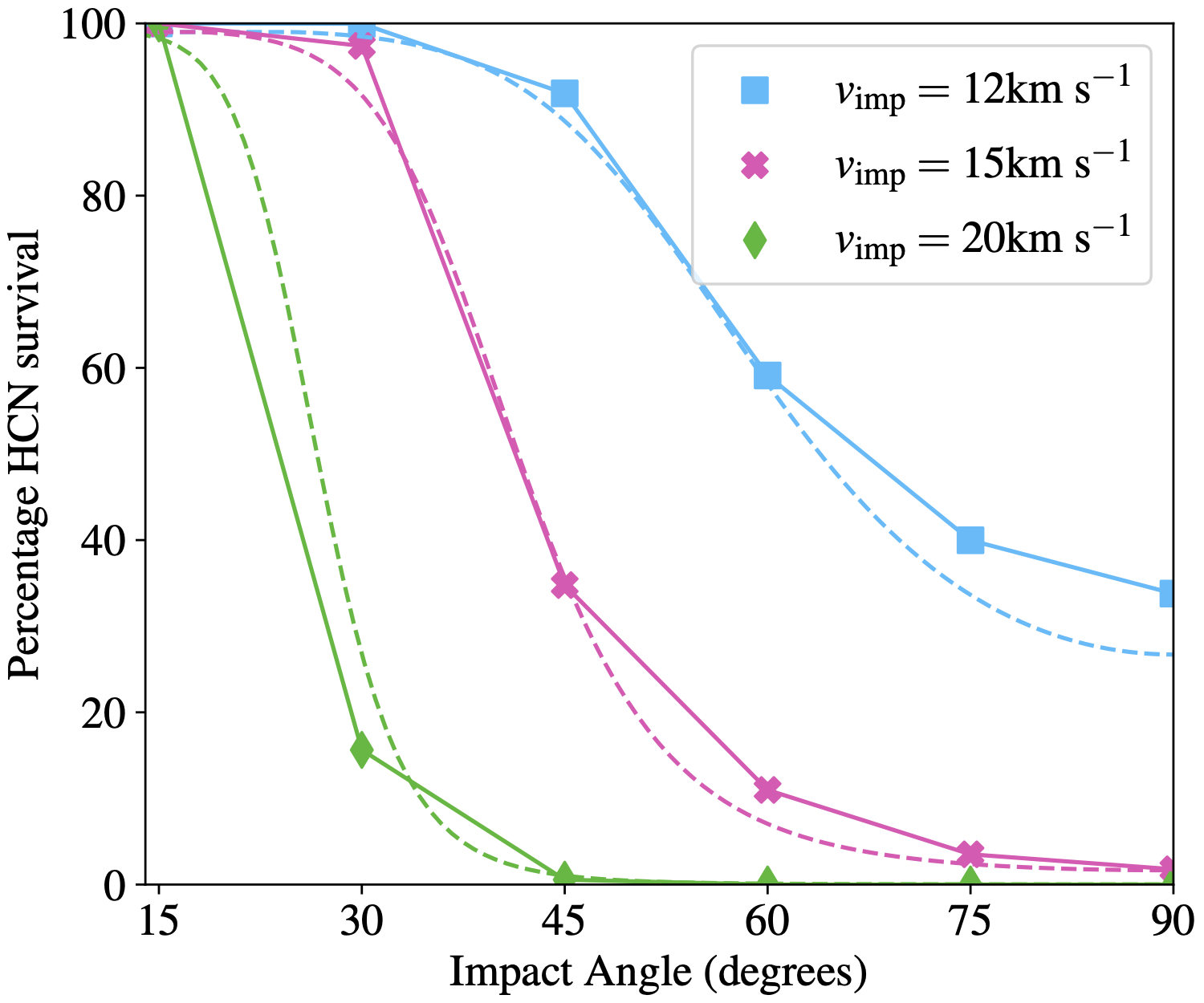}
        \label{subfig:full_kin}
        \end{subfigure}
    \caption{The total surviving percentage of HCN throughout a $1$~km diameter cometary body which survives as a function of impact angle assuming equilibrium chemistry (a) and steady state kinetic chemistry (b).
    The effect of impact velocity is shown by coloured markers as described in the figure legends.
    With decreasing impact angle (and corresponding decreasing temperatures), the survival of HCN increases. 
    The dashed lines show the parametrised survival of HCN as described in Section~\ref{subsec:idealimpact} and given by Equations~\ref{eq:full_param}-\ref{eq:full_param_Gamma}.}
    \label{fig:HCNsurv_angle}
\end{figure*}
In Figure~\ref{fig:HCNsurv_angle} we show the total survival of HCN within a $1$~km comet as a function of impact angle for a range of impact velocities, assuming equilibrium chemistry (Figure~\ref{subfig:full_equil}) and steady state kinetic chemistry (Figure~\ref{subfig:full_kin}).
Largely, it can be seen that reducing the angle of impact increases the survival of HCN by decreasing impact induced temperatures as in Figure~\ref{fig:angleScaling}.
The dashed lines show the full parametrisation of HCN survival as a function of impact velocity, angle, and size as discussed in Section~\ref{subsec:idealimpact}.

Directly comparing our results with those of \citetalias{ToddOberg2020:HCNcomet} presents difficulties due to fundamental differences between our analyses.
By using the parametrised temperature and pressure curves of \citetalias{PierazzoChyba1999:aminosurvival}, \citetalias{ToddOberg2020:HCNcomet} were restricted to considering 6 different radii shells within the comet, leading to low resolution survival data. 
To provide an estimate of the survival of HCN throughout a fiducial impactor, they integrate their radius shell survivals over all impact angles (assuming an impact angle probability distribution as in \citealt{PierazzoMelosh2000a:ObliqueReview}) and over the volume of the comet. 
For their simplified chemical model which is most similar to ours (see Section~\ref{sssec:limit_chem}) they predict an average survival percentage of $0.7$~\%.

With the caveat that the simulation results of \citetalias{PierazzoChyba1999:aminosurvival} and \citetalias{ToddOberg2020:HCNcomet} are assuming that a comet impacts an ocean layer at $20$~\kms, we can attempt to compare our results. 
Using the parametrised temperature profiles presented in their figure~1a, their impacts achieve a mean peak temperature of $\sim 8458.33$~K, which using Equation~\ref{eq:Tmed_from_vimp} suggests an impact velocity of $v_\text{imp} = 16.46$~\kms for our simulation set up. 
Assuming equilibrium chemistry as in the left panel of Figure~\ref{fig:HCNsurv_angle}, an impact at $v_\text{imp} \sim 16$~\kms would see negligible survival at any impact angle, consistent with the results of \citetalias{ToddOberg2020:HCNcomet}.
However, assuming our steady state kinetic chemistry model, we might expect much higher survival for $v_\text{imp} \sim 16.46$~\kms, around $30$~\% on average across all simulated angles.
Thus, our results appear to be much more sensitive to the particular method of calculating \OHconc than the differences between the full and simplified chemical models of \citetalias{ToddOberg2020:HCNcomet}.

A further similarity between our work and \citetalias{ToddOberg2020:HCNcomet} and \citetalias{PierazzoChyba1999:aminosurvival} is that for $\theta \leq 15^\circ$, survival is almost complete for the impact velocities shown in Figure~\ref{fig:HCNsurv_angle}. 
Although our simulations end with the release of the impactor material from the shocked state, the material will remain at elevated temperatures while the vapour cloud continues to expand and evolve. 
Extending the duration of a single test simulation by $\times \sim 1.6$, we see an appreciable reduction in the survival of HCN for an impact with $v_\text{imp} = 12$~\kms and $\theta = 15^\circ$, from $\sim 99$~\% to $\sim 59$~\%.
However, as iSALE is not well-suited for tracking the evolution of vapour, and our icy impactors will be entirely vaporised, our simulations are unlikely to be reliable at longer durations. 
Our results simply represent the survival of HCN within an idealised comet under the impact impulse and further work would be required to understand how prebiotic feedstocks which survive the initial impact evolve while in an expanding vapour cloud.

The large number of tracer particles included in our simulations allow us to have a much finer understanding of the spatial distribution of surviving HCN. 
\begin{figure}
    \centering
    \includegraphics[alt={These provenance plots show the percentage survival of HCN throughout a comet impacting at 12 kilometres a second for three different angles. At an impact angle of 30 degrees, the majority of the body retains 100 percent of its HCN. at the leading edge of the impactor approximately a quarter of the body has all of its HCN destroyed. At an impact angle of 45 degrees, the majority of the HCN is now destroyed, with a small section of the body retaining 100 percent of its HCN towards the outside right edge of the body. Finally, at an impact angle of 60 degrees, the area with high survival is even narrower and towards the top of the impactor, approximately opposite the impact point.}, width=\textwidth]{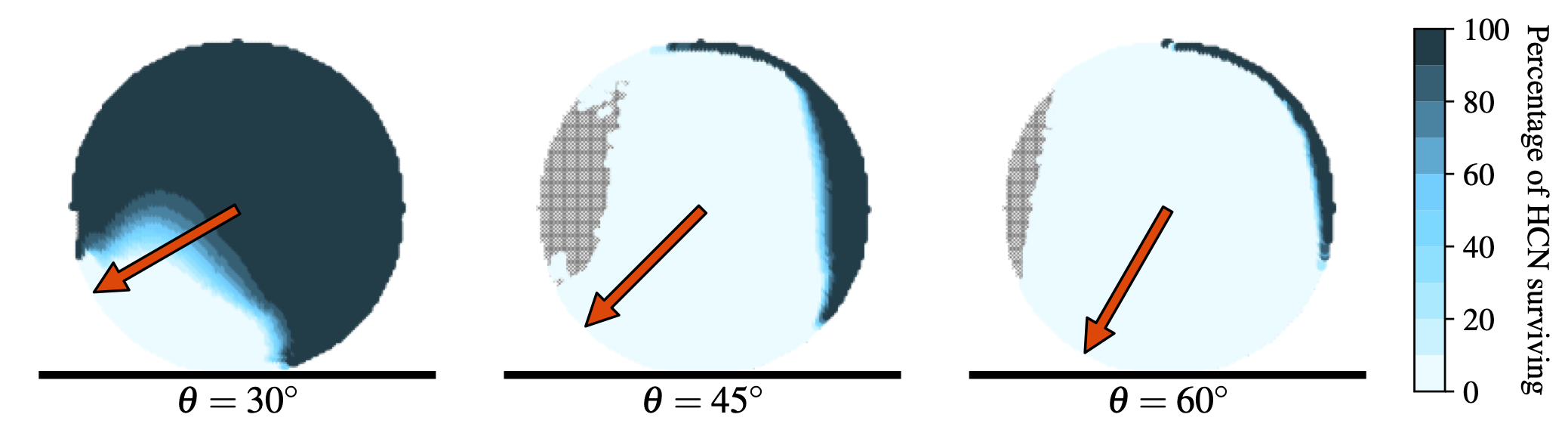}
    \caption{Spatial survival of HCN throughout a comet during a $12$~\kms impact for three impact angles assuming equilibrium chemistry.
    The survival of HCN is calculated for each tracer particle and shown at the tracer's original location within the body, where the colour shows the surviving percentage as described in the colour bar. 
    The grey shaded indicates regions where tracer particles are removed prior to the end of the simulation, we note that these tracers are predominantly contained to regions where we expect zero survival, and are removed after they have been released from the shock and so we do not lose any information about survival through their removal.
    The overlaid red arrow indicates the direction of the comet's motion when it impacts the surface of the planet as shown by the horizontal black line.}
    \label{fig:oblique_spatial_surv}
\end{figure}
Figure~\ref{fig:oblique_spatial_surv} shows the surviving percentage of HCN for each tracer particle (assuming equilibrium chemistry) placed back at their original locations (as in Figure~\ref{fig:HCN_spatial_surv}), for three different impact angles ($\theta = \left[ 30, \, 45, \, 60\right]^\circ$) at $v_\text{imp} = 12$~\kms. 
As in the 2D results shown in Figure~\ref{fig:HCN_spatial_surv}, survival is lowest at the leading edge of the impacting object, and highest towards the opposite edge which will be in a shocked state for the least amount of time.

\section{Discussion} \label{sec:discussion}
This work concludes that in the most optimal scenarios, delivery of HCN is only possible for low velocity, grazing cometary impacts, conditions which may have been difficult to achieve on the early Earth.
In the following, we discuss the limitations of our model, with a focus on whether the conclusion of some limited HCN survival in cometary impacts is robust, and the implications of such survival on the prospects of prebiotic chemistry.
We start by summarising the survival of HCN as a function of cometary impactor parameters, then outline the key simplifications we have made during the course of our modelling, and finally what our results may tell us about the role of comets in prebiotic chemistry.

\subsection{The ideal cometary impact} \label{subsec:idealimpact}
The systematic parameter exploration we have carried out in this work allows us to summarise HCN survival into one, somewhat unwieldy equation for any arbitrary idealised impact  with $v_\text{imp}$, $\theta$, $d$ as follows 
\begin{equation}
    \gamma \left( v_\text{imp}~\text{km s}^{-1}, \, \theta^\circ, \, d~\text{km}\right) = \frac{AB}{1 + \text{exp} \left\{ q \left( v_\text{imp}\,^{1/2} \, \text{sin} \, \theta \,^{0.37} - v_\text{0}\,^{1/2} \right) \right\}} d\,^{\Gamma(v_\text{imp}, \, \theta)}
    \label{eq:full_param}
\end{equation}
where
\begin{equation}
    \Gamma (v_\text{imp}~\text{km s}^{-1}, \, \theta^\circ) =  \left(C \, v_\text{imp} \, \text{sin} \, \theta \,^{0.74} \right)^{E}.
    \label{eq:full_param_Gamma}
\end{equation}
\begin{table}
    \centering
    \begin{tabular}{c|cc}
         & Equilibrium & Kinetic \\
         \hline
         $A$ & $99.64$ & $100.22$ \\
         $q$~(s$^{1/2}$~km$^{-1/2}$) & $7.99$ & $7.53$ \\
         $v_0$~(\kms) & $7.63$ & $11.10$ \\
         $B$~(km $^{-1}$) & $0.97$ & $0.99$ \\
         $C$~(s~km$^{-1}$) & $0.085$ & $0.068$ \\
         $E$ & $5.61$ & $6.63$ \\
    \end{tabular}
    \caption{Parameters for determining the parameterized HCN survival during a comet impact using Equations~\ref{eq:full_param}-\ref{eq:full_param_Gamma} assuming equilibrium and kinetic chemistry models.}
    \label{tab:chem_params}
\end{table}
 Table~\ref{tab:chem_params} summarises the parameters in Equations~\ref{eq:full_param}-\ref{eq:full_param_Gamma} assuming both equilibrium and kinetic chemistry, and the parametrisation is shown as the dashed lines in Figure~\ref{fig:HCNsurv_angle}.

\begin{figure}
    \centering
    \includegraphics[alt={The left panel shows that the percentage survival of HCN increases with decreasing impact angle and decreasing impact velocity. The right panel shows survival of HCN is only possible at impact velocities below approximately 11 kilometres a second, and is approximately constant for all impactor diameters.}, width=\textwidth]{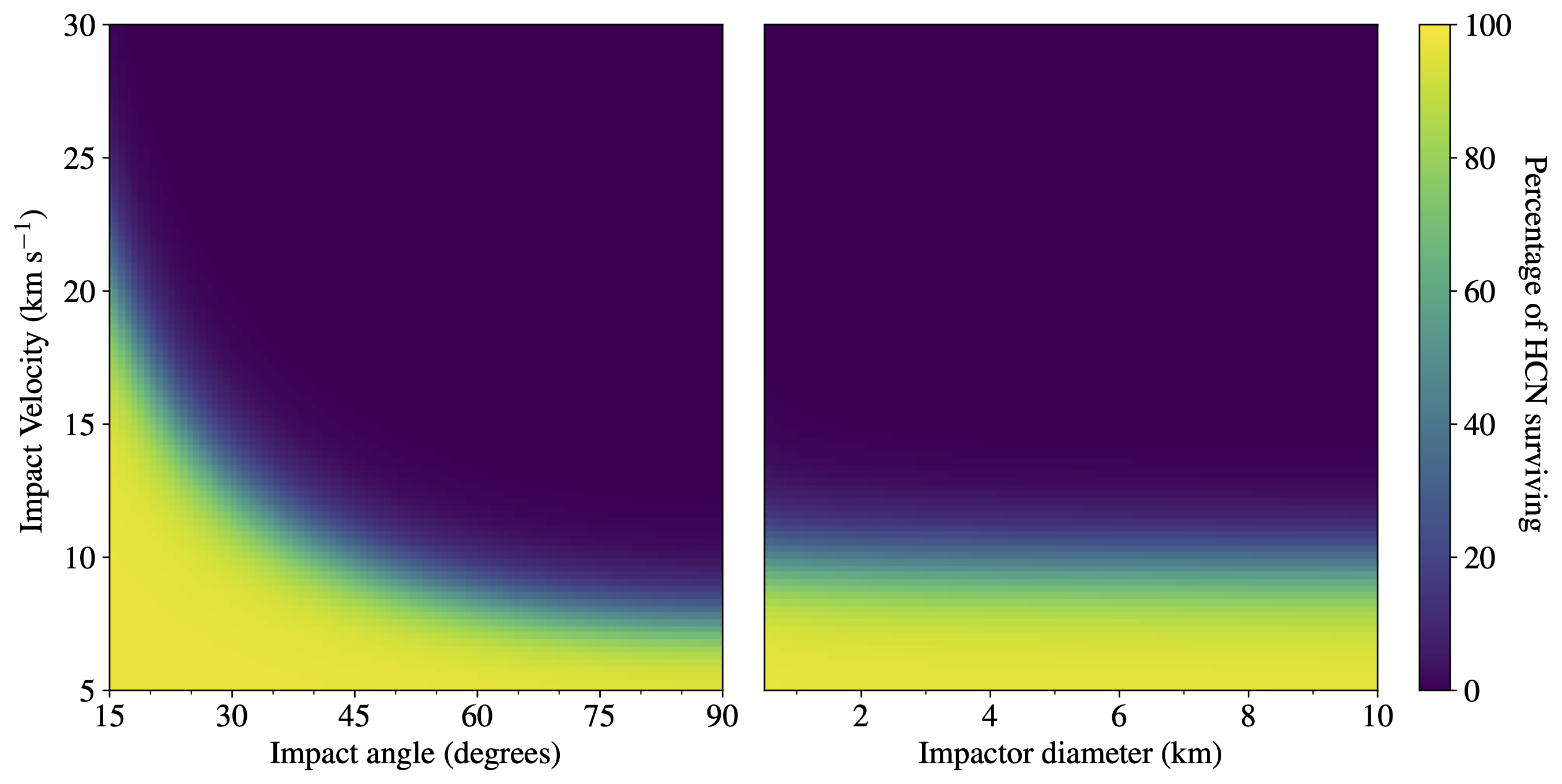}
    \caption{The percentage of HCN surviving impacts as a function of impact velocity and impact angle (left panel) and impact velocity and impact diameter (right panel) as described in the colour bar assuming equilibrium concentrations of OH.
    In the left panel the impactor diameter is kept constant at $1$~km diameter and in the right panel all impacts occur at $45^\circ$.}
    \label{fig:surv_param_space}
\end{figure}
In Figure~\ref{fig:surv_param_space}, we display the full parameter space as described by Equations~\ref{eq:full_param}-\ref{eq:full_param_Gamma} (assuming equilibrium \OHconc), to aid with identification of successful impacts.
The left hand panel shows the survival as a function of impact velocity and impact angle for a $1$~km diameter comet, as before the highest survival rates are seen for the lowest impact velocities and angles. 
The right hand panel shows the survival as a function of impact velocity and impactor diameter at a constant impact angle of $45^\circ$, we again see that the impactor size has a negligible effect on the survival rates, only increasing slightly for the very smallest diameter objects.
Although this figure (and this manuscript in general) neglects to consider the effect of atmospheric passage on the impact parameters at the surface, this is the focus of an upcoming study (Anslow \& McDonald, et al., in prep).

Alternatively, the survival of HCN can be calculated following this relatively simple five step recipe for any $\{v_\text{imp}$, $\theta$, $d\}$ triplet.

\begin{enumerate}
    \item Calculate the median temperature, $T_\text{med}$, experienced by the cometary body from $v_\text{imp}$ as given by Equation~\ref{eq:Tmed_from_vimp} and Figure~\ref{fig:Vel_temp}.
    \item Scale $T_\text{med}$ by the impact angle $\theta$ as in Equation~\ref{eq:Tmean_theta_scale} and Figure~\ref{fig:angleScaling}.
    \item Calculate the equivalent $v_\text{imp, equiv}$ for the scaled $T_\text{med}$ using Equation~\ref{eq:Tmed_from_vimp}.
    \item Calculate the surviving HCN percentage $\gamma$ for $v_\text{imp, equiv}$ using Equation~\ref{eq:surv_vimp}.
    \item Finally, scale $\gamma$ to the impactor diameter, $d$, using Equations~\ref{eq:size_scaling}-\ref{eq:sizescaling_velparam}.
\end{enumerate}

Utilising Equation~\ref{eq:full_param}, and the initial mass of HCN homogeneously distributed throughout the comet, we can estimate how much HCN is delivered during a successful impact.
A single, large ($10$~km), slow ($10$~\kms), grazing ($15^\circ$) impact could deliver as much as $1.15 \times 10^{12}$~kg of HCN. 
Around $78,000$ smaller ($0.5$~km) comets, impacting at $12$~\kms and $45^\circ$ would be needed to deliver an equivalent mass of HCN.

\subsection{Model limitations} \label{ssec:DiscLimitations}
The result that HCN survival is possible in some of our impact simulations is heavily reliant on the specific simplifying assumptions and choices we have made in our modelling, which largely represents best-case scenarios for delivery through impacts. 
Below we outline some of the important considerations which have been neglected in our simple modelling, and the expected impact they have on our results.

\subsubsection{Cometary Parameters} \label{sssec:limit_comet}
\textit{We only consider spherical comets with zero porosity.}
In situ measurements indicate that comets have extremely high porosity, in the range $69-75$~\% \citep{AHearn2005:DeepImpact, Patel2015:cyanosulfidic}, meaning a large proportion of cometary bodies are voids, which we do not model here. 
\cite{PotterCollins2013:AsteroidSurvivability} and \cite{Wunnemann2008:PorousMelt} find that including porosity leads to increased mechanical work and increased temperatures during impacts. 
The results of \cite{Wunnemann2008:PorousMelt} suggest that the increasing temperatures plateau above a porosity of $40$~\%.
Including porosity in our simulations would decrease HCN survival through increased temperatures.

In reality, comets are not spherical and display highly irregular and asymmetric shapes (such as the famed `duck' shape of Comet 67P/ Churyumov–Gerasimenko, e.g. \citealt{Massironi2015:67P}). 
\cite{PotterCollins2013:AsteroidSurvivability} investigate how much of an asteroidal impactor is melted/vaporised during an impact using prolate and oblate shape models.
They find that a lower percentage ($0.2~\%$) of an oblate asteroid remains solid during an impact compared to a spherical asteroid ($3~\%$) or a prolate asteroid ($33~\%$). 
They posit that the changes in survival with shape are due to interactions between the shock wave and the free surfaces of the impactor providing a shielding effect towards the tops of elongated objects.
Although we neglected to consider any shape change caused by atmospheric passage, a deformation model such as pancaking could result in a cylindrical shaped comet \citep{Chyba1993:Tunguska}.
Studies focussed on impacts between spacecraft and asteroids have used more complicated shapes such as cylinders but consistently found that the peak pressures achieved during impact are similar for all shapes \citep[e.g.][]{Kadono2022:ProjShape, Raducan2022:DARTGeometry}.
More complicated shape models which would better describe real cometary shapes, and account for changes in the comet's shape due to atmospheric passage would be computationally prohibitive, but would no doubt show even more complex impact heating distributions, which could significantly affect the survival of HCN within a cometary impactor.

\textit{Our 2D and 3D simulations model pure ice hydrodynamically.}
Our models only consider solid water ice and do not include any representation of cometary dust, which is found to be largely silicate \citep[see][for a review of cometary dust]{LevasseurRegourd2018:CometaryDust}.
This choice simplifies our simulation procedure, allowing us to include only one material and to utilise a hydrodynamic strength model, but does not permit a realistic representation of a cometary body. 
Our use of a hydrodynamic strength model is reasonable with our ice impactors as even at low impact velocities the impactor material does not remain in the solid state. 
Introducing silicate dust grains would necessitate more careful treatment of strength, to fully capture the temperatures achieved during impacts. 
Figure 3 of \citep{Halim2021:MoonBiomarkers} shows how the impact temperatures within a silicate projectile vary by a few hundred Kelvin between purely hydrodynamic simulations and those with strength models at low impact velocities ($v_\text{imp} < 5$~\kms).
Of particular importance to the slowest impact velocities considered in our work ($v_\text{imp} < 10$~\kms) would be considering plastic heating, which would increase the temperature profiles within the impactor \citep{KurosawaGenda2018:PlasticDeformation}, reducing HCN survival for more realistic cometary bodies.

\subsubsection{Chemistry Parameters} \label{sssec:limit_chem}
\textit{We utilise a simple chemical composition and chemical evolution model.}
Comets are known to contain a startling array of volatile rich molecules \citep[e.g. HCN, C$_2$H$_6$, OCS; ][]{MummaCharnley2011:CometChemRev}, and perhaps as much as 5-10~\% of the rocky material within comets is in the form of ammoniated salts \citep{MummaCharnley2019:AmmoniatedSalts}.
For simplicity, we assume that our comets consist solely of H$_2$O and HCN, neglecting to include any additional moleccules. 
This in turn, restricts the complexity of chemistry which we can model.

Our chemical analysis only considers two reactions (Equations~\ref{eq:thermalWaterDecomp}-\ref{eq:HCNdegrad}), similar to the `simplified' chemical network of \citetalias{ToddOberg2020:HCNcomet}.
The alternative `full' chemical model of \citetalias{ToddOberg2020:HCNcomet} contains 31 chemical species and 111 reactions.
Their results suggest that there is only a difference of a factor of a few between the results of the full and simplified models, implying that the survival of HCN is relatively insensitive to the precise details of the chemistry. 

We do however note that since this work (and \citetalias{ToddOberg2020:HCNcomet}), only includes HCN as a nitrogen-carrier, the possibility of HCN synthesis during the impact itself is neglected. 
Considering synthesis would necessitate additional chemistry to be included within the impacting comet, and a careful consideration of the early Earth's atmospheric composition \citep{Parkos2018:EjectaHCN}.

\textit{Finally, we assume that HCN is uniformly distributed throughout the cometary body.}
The chemical inventory of UCAMMs suggest that they formed in the outermost regions of the Solar System \citep{Dartois2013:AntarcticMicrometeorites}, likely from the same reservoir of dusty ice that accumulated to form comets.
Therefore, we may expect localised regions of a cometary body to contain more compact dusty inclusions which display distinct chemical compositions. 
Understanding the fate of this dust within the body of a comet is beyond the scope of this work but provides a further complication necessary to consider the full capability of comets to deliver organics (see Section~\ref{sssec:limit_comet}).
Flyby observations of 9P/Tempel 1 and 103P/Hartley 2 have shown distinct regions of activity coupled with jets of CO$_2$ and H$_2$O that indicate compositional variations consistent with comet formation theories \citep[e.g.][]{Farnham2007:9PTempel1, MummaCharnley2011:CometChemRev, BruckSyal2013:103PHartley}.
Observations of the dynamically new comet C/2012 S1 (ISON) showed varying abundances of HCN as the comet approached perihelion, indicative of variations in the abundance of HCN within the body of the comet \citep{DiSanti2016:CometS1, DelloRusso2016:CometS1}.
Further, modelling of the full chemical composition of cometary nuclei through solid-phase chemical kinetics find that HCN should have a relatively consistent abundance throughout a few metres of cometary body \citep{Garrod2019:CometIceChem}, highlighting the continued modelling required to accurately capture the composition of comets.

Knowing the true location of HCN within a comet body is required in order to assess its survival during a cometary impact. 
If a large concentration of HCN is localised towards the trailing edge of the impactor, this would show a much larger survival fraction compared to at the leading edge. 
Applying our impact temperature-pressure profiles to more detailed models of cometary composition would allow us to put more realistic constraints on HCN survival.

\subsubsection{Early Earth} \label{sssec:limit_earth}
\textit{We do not consider evolution of the comet as it passes through the atmosphere.}
We chose to neglect the evolution of the body's size and shape during the atmospheric passage since the size of the objects we consider will mostly be unaffected by the passage due to their size.
The smallest comets we consider in this work with a diameter of $500$~m are on the edge of what size comets are predicted to reach the surface of the Earth without undergoing fragmentation \citep[e.g.][]{ArtemievaShuvalov2001:Fragments, Anslow2024:OverlappingCraters}.
Thus, we believe that the range of impact sizes and velocities we consider in this work should cover the possible range of objects which will survive atmospheric passage and arrive at the surface of the Earth. 
Due to the short period of time a comet is traversing the atmosphere, it is unlikely that any atmosphere-induced chemistry will be able to propagate through the body and will be restricted to the leading edge, which is essentially destroyed in all of our simulations. 
Although there is a large amount of uncertainty in the surface density of the Hadean Earth atmosphere, we thus may naively expect to not see much difference in our results by including atmospheric passage induced chemistry.

\textit{We assume the impact occurs on solid ground.}
The cyanosulfidic pathway to life, as described in this work, relies on HCN being delivered to an isolated aqueous environment, thus by necessity an impact successfully delivering HCN needs to occur on solid ground.
The timing, and extent, of continental crust formation is disputed \cite[e.g.][]{GuoKorenaga2020:ContinentCrust, Pujol2013:Crust, Campbell2003:Continent}, but currently covers $\sim 40$~\% of the surface of the Earth.
If impacts occur isotropically over the surface of the Earth, then only $40$~\% of impacts have the potential to `successfully' delivering HCN as defined in this work.
The majority of impacts would occur into an ocean layer, which presumably motivated the modelling choices of \citetalias{PierazzoChyba1999:aminosurvival}.
Impacts into water will achieve lower temperatures, which may subsequently lead to increased levels of HCN survival as seen in \citetalias{PierazzoChyba1999:aminosurvival} and \citetalias{ToddOberg2020:HCNcomet}.
However, delivering to oceans only exacerbates the difficulties in concentrating HCN in usable reservoirs, as discussed further below.
Thus, for the prospects of prebiotic chemistry, the extreme temperatures involved with impacts on the Earth's surface may be offset by the dilution of HCN within the ocean.

\subsection{Implications for prebiotic chemistry} \label{ssec:Implications}
Our results highlight that in a narrow range of impact events (small, slow and oblique), HCN could be delivered to the early Earth in our idealised parametrised model.
The HCN-enriched water vapour produced by such successful impact delivery events will eventually cool and condense, potentially accumulating in small isolated ponds. 
As envisioned by \cite{Patel2015:cyanosulfidic} and \cite{Sasselov2020:Origins} for the cyanosulfidic pathway to life, episodic drying events could then store the HCN as ferrocyanide salts, allowing the accumulation of larger concentrations of HCN. 

This work has focussed on the implications of comet delivery on the cyanosulfidic scenario, but HCN has been widely implicated as an essential feedstock for other prebiotic chemistries \citep{Sutherland2016:BlueOrigins}.
HCN reacting with aldehydes and ammonia can form amino acids like glycine, alanine and serine through Strecker synthesis \citep[e.g.][]{Moutou1995:Strecker, Singh2022:Strecker}.
The production of adenine by HCN in aqueous solution \citep{Oro1960:AdenineSynthesis} has been suggested as the fundamental step in producing RNA precursors that formed the `RNA world' \citep[e.g.][]{Orgel2004:RNAworld, Das2019:RNAworld, Muller2022:RNApeptideWorld}.
Formamide has been suggested as an alternative method to store HCN, and as a stepping stone towards producing other prebiotic molecules such as purine \citep{Sanchez1966:Purine} and wider prebiotic chemistries \citep{Saladino2015:MeteoritesNucleosides, Saladino2016:MeteoritesFormamide}. 
Thus, the successful delivery of HCN by cometary impacts not only benefits the cyanosulfidic pathway to life, but many others, extending the results of this study to the wider origins community.

However, if comets contributed significantly to the onset of life in this way, it was likely through a chance delivery event that was followed by many steps not modelled in this work.
Thus, although it may be possible for HCN to be delivered to the early Earth as outlined above, this does not necessarily lead to the direct onset of a chemical pathway to life, we outline some of the further complications to this picture below. 

\subsubsection{Stockpiling HCN} \label{sssec:stockpiling}
Even if HCN can survive the impact process, it may not result in HCN concentration within a favourable environment for prebiotic chemistry to proceed.

As comets contain a large fraction of water ice and other volatile species, we would expect that the majority of the impactor material will expand out of the impact crater, in the direction of travel as a vapour plume.
The highest energies and temperatures achieved during the impact process occur at the leading edge of the body in the heart of the forming crater, so the material most likely to retain HCN is also likely to be within the rapidly expanding vapour cloud produced towards the rear of the projectile. 
Within our simulations we see a strong obliquity dependence on the number of tracer particles which are removed from our oblique simulations, which is shown by grey shaded regions in Figure~\ref{fig:oblique_spatial_surv}.
These tracers are removed due to drifting too far from the material they are supposed to be tracing, highlighting the rapid movement of the impactor material away from the impact crater.
The greater number of particles being removed from the $\theta = 45^\circ$ simulation may indicate that these impact angles are most efficient at funnelling generated vapour out of the impact crater.

Vapour is produced after the impactor material has been released from its shocked state and is able to expand outwards, and thus preferentially occurs at the edge of the forming impact crater and the rear of the projectile.
Given that our simulations only extend to the end of rarefraction, and that iSALE is not ideally suited to tracking vapour once it has been generated, our results only describe the very beginning of vapour formation.
For oblique impacts, \cite{Schultz1996:ObliqueVaporization} find that the vapour cloud produced may decouple from the crater formation process earlier than in vertical impacts and move ballistically downrange. 
Thus, detailed modelling of the produced vapour and its subsequent evolution is of great importance to evaluating the ultimate concentration of HCN, but is beyond the scope of this manuscript. 

If HCN can be be accumulated within an isolated pond, in aqueous solution HCN will undergo hydrolysis ultimately producing formic acid and preventing chemical pathways from proceeding. 
It has been proposed that storing the HCN in a more stable form such as ferrocyanide, from which the cyanide may later be liberated (e.g. through thermal processing) could circumvent the issues of HCN destruction \citep[e.g.][]{Todd2024:FerrocyanideEnvirons} and to some extent the difficulties in concentration. 

However, millimolar concentrations of ferrocyanide itself in aqueous solution have a half life on the order of hours under UV radiation \citep{Todd2022:Ferrocyanide}, suggesting shallow ponds of water could similarly struggle to stockpile ferrocyanide.
\cite{Sasselov2020:Origins} has instead suggested that precipitating the ferrocyanide in salts which can accumulate through repeated evaporative cycles, could provide the solution.
The lifetime of ferrocyanide in salt form is not well constrained and further experiments would be required to elucidate its storage capabilities. 

Even if sufficient stockpiles of ferrocyanide salts are able to accumulate, additional processing is required to free the cyanide through energy sources such as igneous intrusions, volcanism, or impacts \citep{Sasselov2020:Origins}.
\cite{Anslow2024:OverlappingCraters} investigate the scenario where a second impactor provides energy to a pond containing a stockpile of ferrocyanide by estimating the number of overlapping cratering events within a range of possible ferrocyanide lifetimes. 
They find that it is incredibly unlikely that double impactors could produce environments required for prebiotic chemistry, unless the lifetimes of ferrocyanide salts are unexpectedly long, motivating further studies into the later stages required to build habitable conditions. 

Thus, even though we find that a truly fortuitous cometary impact could deliver HCN, its delivery and concentration must outpace its destruction, and there remains a suite of outstanding questions related to the role cometary impacts may play in the onset of prebiotic chemistry.

\subsubsection{Alternative HCN accumulation} \label{sssec:alternativeHCN}
Although we neglected to consider the interaction of the impacting comets with the atmosphere, this process could be a vital source for organic molecules.
\cite{SainsburyMartinez2024:ImpactsTerrestrialExo} model the evolution of mass deposited in the atmosphere of a tidally-locked terrestrial planet (analogous to the Trappist planets) by cometary ablation using a 3D global circulation model. 
They find that deposited water vapour is rapidly transported both vertically and zonally altering the composition and thermal structure of the atmosphere. 
The atmospheric rotation of the Earth and its circulation patterns may lead to more homogeneous latitudinal mixing but slower vertical mixing \citep[e.g.][]{Showman2013:AtmosphericCirculation}.
Thus, delivering HCN to the surface of the Earth through rainout would depend on the atmospheric depth at which the break up occurred and the ability for the HCN to survive any further chemical reactions in the poorly constrained atmosphere of the early Earth.

Alternatively, the high energy process of an impacting body interacting with the planetary atmosphere may produce important prebiotic chemicals.
Laser-driven plasma experiments suggest that during an impact formamide may participate in chemistry to produce nucleobases \citep{Ferus2015:Formamide} and that RNA nucleobases may be formed in Miller-Urey-like atmospheres \citep{Ferus2017:MillerUreyNucleobases}.
The shock wave produced by an impactor passing through the atmosphere can also initialise the production of amino acids and other organic molecules through `post-impact quench synthesis' \citep[e.g.][]{BarakBarNun1975:QuenchSynthesis, Parkos2018:EjectaHCN}.

Impacts of asteroidal meteorites may also provide an important source of HCN. 
Large, highly reduced and differentiated bodies whose impacts are capable of vaporising oceans can produce transient steam atmospheres where photochemistry can generate HCN and other prebiotic nitriles \citep{Zahnle2020:ImpactAtmos, Wogan2023:ImpactAtmos}.
Interactions between an atmosphere rich in HCN and an aqueous environment in contact with iron-rich bedrock could produce ferrocyanide for storage \citep{Sasselov2020:Origins} and later participation in prebiotic chemistry.
Key prebiotic feedstocks, including HCN, can also be abundantly produced in shallow hydrothermal systems \citep{RimmerShorttle2019:CNHydroVents, RimmerShorttle2024:SurfaceHydrothermals}, given the volcanic activity present on the early Earth, hydrothermal vents could provide a method to produce environments enriched in prebiotic feedstocks.

Finally, much smaller extraterrestrial sources of HCN and other organic molecules could be key to the concentration of prebiotically important chemicals. 
Interplanetary dust particles may be an important source of carbon, nitrogen, and hydrogen early in the terrestrial planet's histories \citep{AndersOwen1977:TerrestrialVolatiles}, and UCAMMs currently constitute a substantial amount of the carbon delivered to the Earth \citep{Rojas2021:MicrometeoriteFlux}. 
Falling interplanetary dust can be accumulated in regions favourable to prebiotic chemistry through sedimentary processes driven by glaciers \citep{Walton2024:IDPs}. 
Thus, further holistic modelling of the chemical processes driven by impacts are required to fully understand the role of comets in accumulating organic molecules.

\subsubsection{Delivery of Other Molecules} \label{sssec:OtherMolecules}
As interesting as HCN is in its own right, it also acts as a useful proxy molecule for more general impact delivery, since its CN triple-bond means the molecule is far more resistant to impact thermolysis than almost any other prebiotically relevant molecule. 
Cyano radicals (CN) even exist in the atmosphere of the Sun \citep{Bauscchlicher1987:CNDissociation}. 
The fraction of prebiotic molecules such as amino acids, sugars, or nitrogen-containing rings (e.g., pyrimidines or purines) that can survive an impact process will be significantly lower than the fraction of HCN that would survive. 
Therefore, we have good reason to believe that in our impact simulations (Section~\ref{subsec:impact_3D}), where HCN is destroyed, other more fragile prebiotic molecules will also be destroyed.
However, where HCN survives an impact process, we will have to undertake further work to assess the survival of other molecules such as amino acids, nucleobases, sugars \citep[see][]{Zellner2020:Glycolaldehyde}, or fatty acids.
Of particular interest is cyanoacetylene (HC$_3$N), a strong molecule (although not as resilient as HCN), which is essential for many prebiotic chemical scenarios \citep{Benner2020:RNA, Green2021:UVPhotochem, Orgel2002:Cyanoacetylene, Muller2022:RNApeptideWorld}. 
Comets are rich in HC$_3$N \citep{MummaCharnley2011:CometChemRev, Oberg2015:CometPPD}, and so far there is no plausible way known to obtain prebiotically relevant concentrations of HC$_3$N other than successful cometary delivery, though there are some hypothetical alternative sources \citep{RimmerShorttle2019:CNHydroVents, RimmerShorttle2024:SurfaceHydrothermals, Wogan2023:ImpactAtmos}.
Thus, the impact conditions where HCN survives should be further explored to firstly consider the survival of HC$_3$N, and then be extended to other more fragile molecules closer to those used by biology (amino acids, aminonitriles, etc.).

\subsubsection{Delivery by Other Bodies} \label{sssec:OtherDelivery}
Although this work has exclusively focussed on the delivery of prebiotic feedstocks through cometary impacts, asteroids are more likely to impact the Earth than comets \citep{YeomansChamberlin2013:Fluxrates}.
The recent identification of large amounts of amino acids. and other volatile molecules, in material returned from the near-Earth asteroid (101955) Bennu \citep{Glavin2025:Bennu, McCoy2025:Bennu}, coupled with surviving amino acids within meteorites \citep[e.g. Murchison;][]{SchimttKopplin2010:Murchison}, has highlighted the importance of also considering delivery of prebiotic feedstocks by asteroids. 
Although we do not simulate asteroid impacts, we can draw some very preliminary conclusions from the work presented here.

Occupying closer orbital distances to the Earth than comets, the average impact velocity of a near-Earth object ($v_\text{imp, NEO} = 18$~\kms; \citealt{Chesley2019:SyntheticImpactors}) is lower than that of either short or long period comets \citep{HughesWilliams2000}.
As shown in Section~\ref{sssec:ImpactVelocity}, lower impact velocities improve the chances of HCN survival, especially when coupled with oblique impact angles.
In Section~\ref{ssubsec:2D_size}, we find that survival of HCN increases with decreasing size. 
The size of body which can impact onto the surface of the early Earth is mediated by passage through the atmosphere \citep{Anslow2025:atmosentry}, smaller asteroids will be able to survive this process due to higher densities and material strengths \citep{Carry2012:AsteroidDensities, Chyba1993:Tunguska}.
Extrapolating just from our cometary simulations, we might then expect HCN survival to be heightened in asteroids.

However, as asteroids contain a smaller water fraction, and higher quantities of silicate-rich dusty and rocky material, impact simulations investigating the prospects for delivering HCN via asteroids, would need much more careful treatment of strength and porosity, both factors which lead to an increase in the temperatures induced by the impact (see Section~\ref{sssec:limit_comet}).
Such an increase in the temperatures involved with asteroidal impacts may then tamper the benefits of smaller and slower impactors. 
Continuing studies of the Bennu sample return material, and further computational modelling, will increase our understanding of the delivery of prebiotic material to the early Earth.

\subsubsection{Delivery to Other Bodies} \label{sssec:OtherBodies}
Finally, we can consider what our updated impact simulations may mean for the prospects of delivering organic molecules to other celestial bodies.

The largest parameter which affects the survival rate of HCN within our impact simulations is the impact velocity, whose lower bound is mandated by the escape velocity of the Earth and the efficiency of aerobraking within the Earth's atmosphere. 
For smaller Solar System bodies such as the Moon, or Mars, the minimum impact velocity is greatly reduced ($v_\text{esc, Moon} \sim 2.4$~\kms and $v_\text{esc, Mars} \sim 5$~\kms), which increases the possibility of cometary impacts being able to successfully deliver HCN or other prebiotically relevant molecules.

Moving first to our closest neighbour, a number of studies have investigated the prospects of delivering organic material from the Earth to the Moon in the form of terrestrial meteorites \citep[e.g.,][]{Armstrong2002:TerranMeteor, Crawford2008:TerrestMeteorites}.
\cite{Halim2021:MoonBiomarkers} present iSALE-3D simulations investigating the survival of biomarker amino acids within terrestrial meteorites impacting the Moon. 
Alongside surviving amino acids, they identify that under certain impact conditions (low velocities and high porosity impact sites) lycophyte megaspores could survive ejection from the surface of the Earth, and subsequent impact onto the surface of the Moon.
Thus, deeply buried terrestrial meteorites on the Moon may provide key information about the biological history of the Earth.

The amino acid survival analysis of \citetalias{PierazzoChyba1999:aminosurvival} has been extended to impacts on Mars \citep{PierazzoChyba1999:Mars, PierazzoChyba2003:MarsOblique, PierazzoChyba2006:PlanetaryDelivery}, finding substantial amino acid survival in cometary impacts, but negligible survival for asteroidal impacts. 
Although the lower escape velocity of Mars permits lower velocity impacts, it also can allow significant amounts of impact ejecta to be liberated from the planet post-impact. 
More oblique impacts cause more impactor material to be ejected.
However, evidence of extensive strata collection in Mars' Gale crater could indicate that the deposition of ferrocyanide salts as discussed above could occur efficiently on Mars \citep{Sasselov2020:Origins}.

Further out in the Solar System, impacts have also been explored as a mechanism for delivering volatile elements and molecules to Ceres \citep{Bowling2020:CeresImpacts}, Europa \citep{PierazzoChyba2002:EuropaDelivery}, and Titan \citep{Hedgepeth2022:TitanMelt}.
The wide array of potentially habitable environments throughout the Solar System would necessitate further dedicated, finer-resolution impact simulations to elucidate the role of cometary delivery across the Solar System.

The possibility of cometary delivery in exoplanetary systems can also be considered. 
\cite{Anslow2023:ExoplanetDelivery} considers the prospects of delivery to chains of  co-planar, equal mass planets \citep[akin to the Kepler `peas-in-a-pod' systems;][]{Weiss2018:PeasInAPod}.
They find that impact velocities around M-dwarf stars like TRAPPIST-1 are significantly higher than those around F- and G-type stars even for low mass planets. 
\cite{Claringbold2023:JWSTbiosigs} suggest that HCN at the ppm level may be detected in temperate rocky exoplanets, like TRAPPIST-1e, with a few transits of observations with JWST. 
Thus, if such an atmospheric concentration of HCN were to be observed within the TRAPPIST-1 planets, we would likely be able to say that this was not delivered to the planet through cometary impacts. 

\section{Conclusions} \label{sec:conc}
The role that comets may play in the origins of life through the delivery of chemical feedstocks has been of interest to the academic community for over a century. 
Extensive laboratory work investigating the chemical reactions which may lead to prebiotic chemistry has identified HCN as a key chemical feedstock whose presence may facilitate the formation of chemical components of life.

This work revisits the idea of cometary delivery through new high-resolution impact simulations of an idealised comet using the shock physics codes iSALE-2D and iSALE-3D.
These new simulations represent a significant improvement over previous numerical studies investigating the survival of HCN by including dedicated 3D oblique simulations and increasing the number of Lagrangian tracer particles included by several orders of magnitude.

Using temperature and pressure profiles recorded by the Lagrangian tracers, we investigated the survivability of HCN during cometary impacts by modelling the OH radical driven degradation of HCN, assuming both equilibrium and steady state kinetic chemistry as a function of impact velocity (Figure~\ref{fig:vel_HCNsurv}), impactor size (Figure~\ref{fig:SizeSurvival}), and impact angle (Figure~\ref{fig:HCNsurv_angle}). 
We find survival drastically decreases with increasing impact velocity and angle, and mildly decreases with increasing comet diameter. 
For both equilibrium and steady state chemistries, highly oblique, grazing impacts are the most suited to delivering HCN. 
Assuming equilibrium chemistry, survival at impact angles $> 15^\circ$ is only possible at impact velocities below the escape velocity of the Earth, thus requiring extreme aerobraking within the atmosphere prior to impact.

In Equations~\ref{eq:full_param} and \ref{eq:full_param_Gamma} we provide a full parametrisation of HCN survival, incorporating the three scaling laws for the parameters considered in this work to produce expressions describing an upper limit on survival.
Figure~\ref{fig:surv_param_space} summarises this parametrisation and emphasises the extreme low velocities and oblique impact angles required in order to deliver intact HCN in our simulations. 

Given the particular resilience of HCN to thermal degradation, and the idealised nature of our simulations, our results suggest difficulty in delivering other prebiotic feedstocks through cometary impacts.
The new impact simulations presented in this work should provide improved understanding of the temperatures and pressures prebiotic molecules must endure to better inform future studies considering more realistic chemical modelling.

Although our results highlight that the delivery of HCN by comets is possible in some limited regions of impactor parameter space, we reiterate that this alone is not sufficient for the onset of prebiotic chemistry and continuing studies into the environments in which the pathway to life can proceed are required. 
While our simulations represent a significant improvement over previous numerical work investigating the survival of HCN, the remaining simplicity of our models requires that our results represent upper limits on survival that should inform and motivate further numerical modelling of the evolution of prebiotic feedstocks within comets impacting on the early Earth.

\section*{Data Availability}
Data will be made available upon reasonable request.

\section*{Acknowledgements}
We thank the anonymous reviewers for their careful consideration of this work and their insightful comments which have improved this manuscript.
We gratefully acknowledge the developers of iSALE-2D and iSALE-3D, including Gareth Collins, Kai W\"{u}nnemann, Dirk Elbeshausen, Tom Davison, Boris Ivanov and Jay Melosh.
This work was performed using resources provided by the Cambridge Service for Data Driven Discovery (CSD3) operated by the University of Cambridge Research Computing Service (\url{www.csd3.cam.ac.uk}), provided by Dell EMC and Intel using Tier-2 funding from the Engineering and Physical Sciences Research Council (capital grant EP/T022159/1), and DiRAC funding from the Science and Technology Facilities Council (\url{www.dirac.ac.uk}).

CHM gratefully acknowledges the Leverhulme Centre for Life in the Universe at the University of Cambridge who have funded this work through Joint Collaborations Research Project Grant GAG/382.
RJA acknowledges the Science and Technology Facilities Council (STFC) for a PhD studentship.
ASPR is grateful for support from Trinity College, University of Cambridge.
This work was additionally funded by the Leverhulme Centre for Life in the Universe Joint Collaborations Research Project Grant G112026, Project KKZA/237.

\bibliographystyle{elsarticle-harv} 
\bibliography{bibliography}

\end{document}